\documentclass[a4paper]{article}
\usepackage{geometry}
\usepackage[utf8]{inputenc}
\usepackage[T1]{fontenc}

\usepackage{moreverb}

\usepackage{graphicx}
\usepackage{xcolor}

\usepackage{amsmath}

\usepackage{algorithm}
\usepackage{algpseudocode}

\algblockdefx{When}{EndWhen}[1]{\textbf{when} #1 \textbf{do}}{\textbf{end when}}
\algrenewcommand\alglinenumber[1]{\small #1:}

\usepackage[caption=false,font=footnotesize]{subfig}

\usepackage{url}

\usepackage{multirow}

\title{%
Network Bandwidth Variation-Adapted State Transfer for Geo-Replicated State Machines and its Application to Dynamic Replica Replacement%
\thanks{A preliminary version of this paper appeared in the ninth International Symposium proceedings on Computing and Networking (CANDAR 2021) \cite{Chiba2021a}.}%
}

\author{Tairi Chiba, Ren Ohmura, and Junya Nakamura\thanks{Corresponding author: junya[at]imc.tut.ac.jp}}

\date{Toyohashi University of Technology, Japan}

\begin{document}

\maketitle

\begin{abstract}
This paper proposes a new state transfer method for geographic state machine replication (SMR) that dynamically allocates the state to be transferred among replicas according to changes in communication bandwidths.
SMR improves fault tolerance by replicating a service to multiple replicas.
When a replica is newly added or recovered from a failure, the other replicas transfer the current state of the service to it.
However, in geographic SMR, the communication bandwidths of replicas are different and constantly changing.
Therefore, existing state transfer methods cannot fully utilize the available bandwidth, and their state transfer time increases.
To overcome this problem, our method divides the state into multiple chunks and assigns them to replicas based on each replica's bandwidth so that the broader a replica's bandwidth is, the more chunks it transfers.
The proposed method also updates the chunk assignment of each replica dynamically based on the currently estimated bandwidth.
The performance evaluation on Amazon EC2 shows that the proposed method reduces the state transfer time by up to 47\% compared to the existing one.
In addition, we apply the proposed method to dynamic replacement of replicas, which can mitigate latency degradation caused by network trouble, and evaluate how fast the method can relocate a replica.
\end{abstract}

\section{Introduction}
\label{sec:introduction}

\emph{State Machine Replication (SMR)} \cite{Schneider1990,Distler2021} is a method that improves the fault tolerance of a service.
SMR replicates a service to multiple servers, called \emph{replicas}, which agree on the order of request processing among them to maintain the same state of each replica.
This method allows replicas to continue the service even if some of them fail, and several SMR protocols have been proposed in previous studies \cite{Bessani2014,Castro1999,Castro2002,Junqueira2011,Kotla2010,Lamport1998,Moniz2011,Nakamura2014,Nakamura2014a,Ongaro2014,Sousa2012}.
Furthermore, performing SMR with multiple geographically separated replicas allows the service to resist large-scale disasters, such as earthquakes.
Such a method is called \emph{geographic SMR}\cite{Amir2010,Berger2019,Coelho2018,Coelho2021,Eischer2018,Eischer2020,Eischer2021,Enes2020,Enes2021,Liu2017,Mao2008,Nawab2018,Ngo2020,Sousa2015,Xu2019,Yan2020,Zhao2018}.

In SMR, a replica that is newly added to replication or is recovered from faults is called a \emph{recovery replica} and retrieves the latest state of the service from the others, called \emph{transfer replicas}, to keep its state the same as that of the others.
This process, called \emph{state transfer}, is important because we cannot avoid faults of replicas in a long-run replication, and this process allows SMR to handle such faults without stopping the entire system.
Schneider introduced a basic state transfer method in which a recovery replica obtains the whole state from a single transfer replica \cite{Schneider1990}.
After that, Bessani et al.~proposed the Collaborative State Transfer (CST) protocol \cite{Bessani2013}, which can mitigate the decrease in the request processing performance during a state transfer. 
In this method, each transfer replica sends an equally sized part of the current state to a recovery replica to reduce the state transfer time.
These existing state transfer methods assume that replicas are deployed in the same data center, where the replicas are connected through a high-speed and stable LAN.

\begin{figure}[t]
\centering
    \subfloat[North Virginia (Worldwide Group)]{
        \includegraphics[width=70mm]{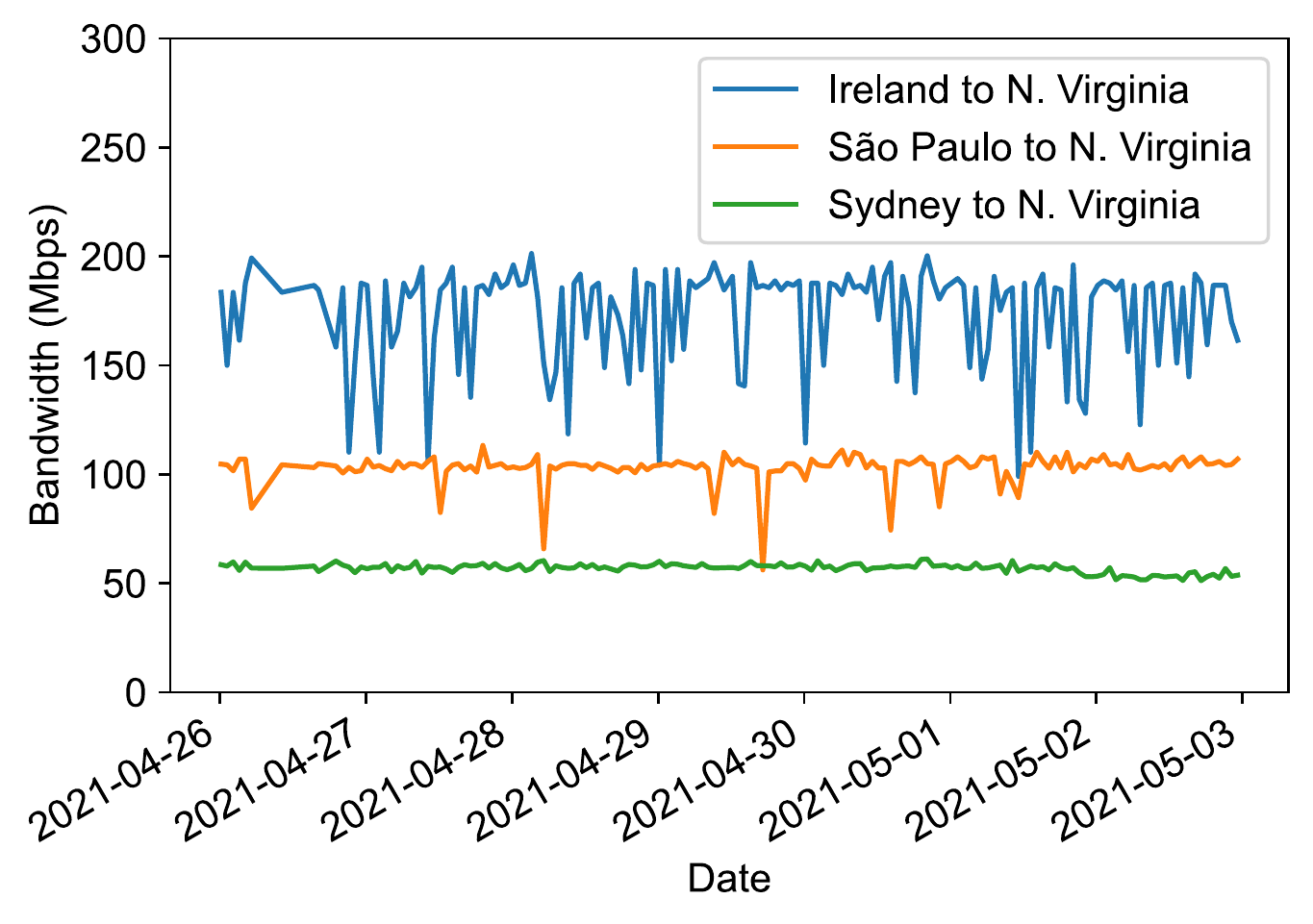}
        \label{fig:bandwidth-change-group-a-virginia}
    }
    \subfloat[London (European Group)]{
        \includegraphics[width=70mm]{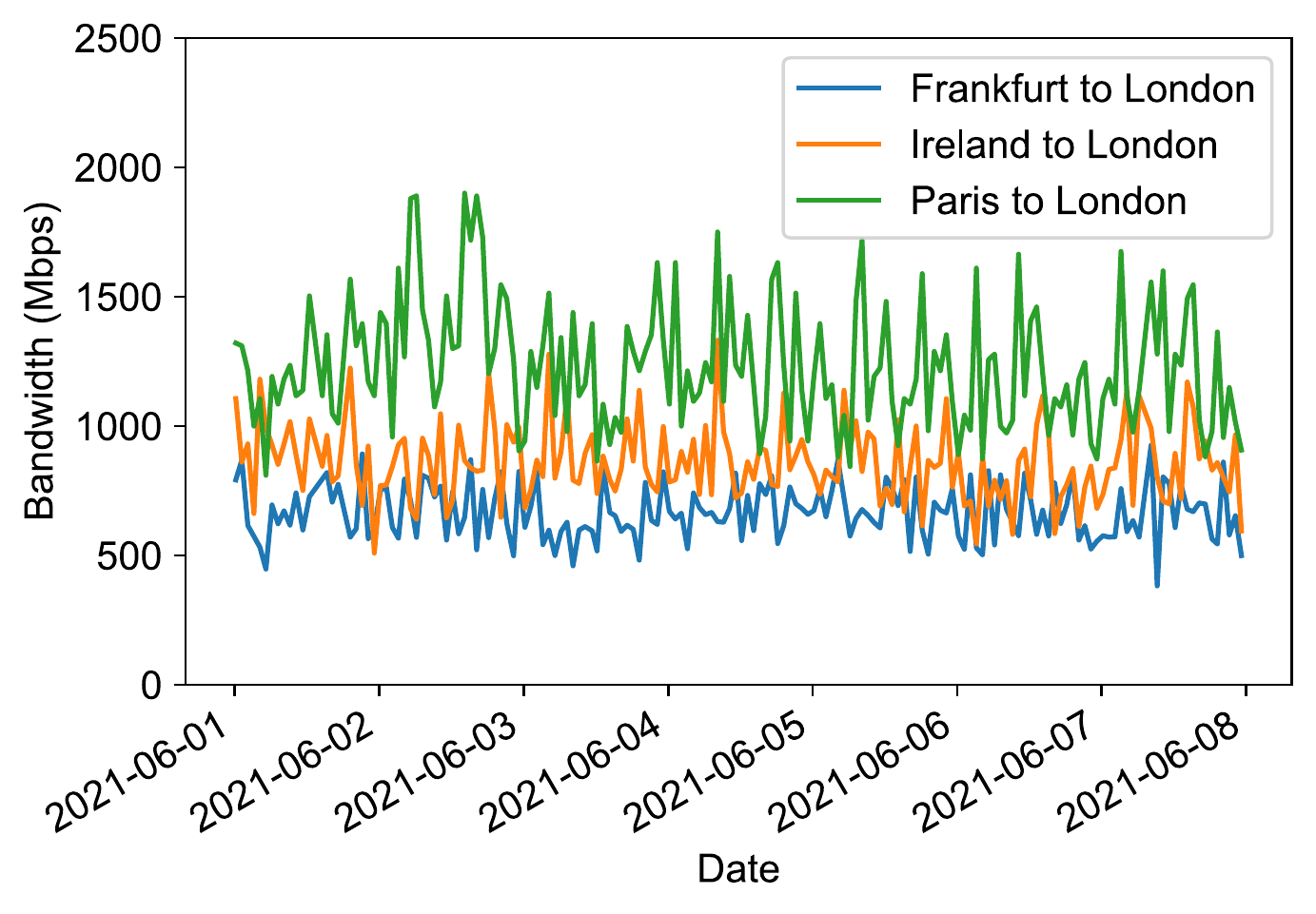}
        \label{fig:bandwidth-change-group-b-london}
    }
    \caption{Time variation of communication bandwidth in Amazon EC2 regions.}
    \label{fig:bandwidth-change}
\end{figure}

\begin{figure}[b]
\centering
    \subfloat[Worldwide Group]{
        \includegraphics[width=75mm]{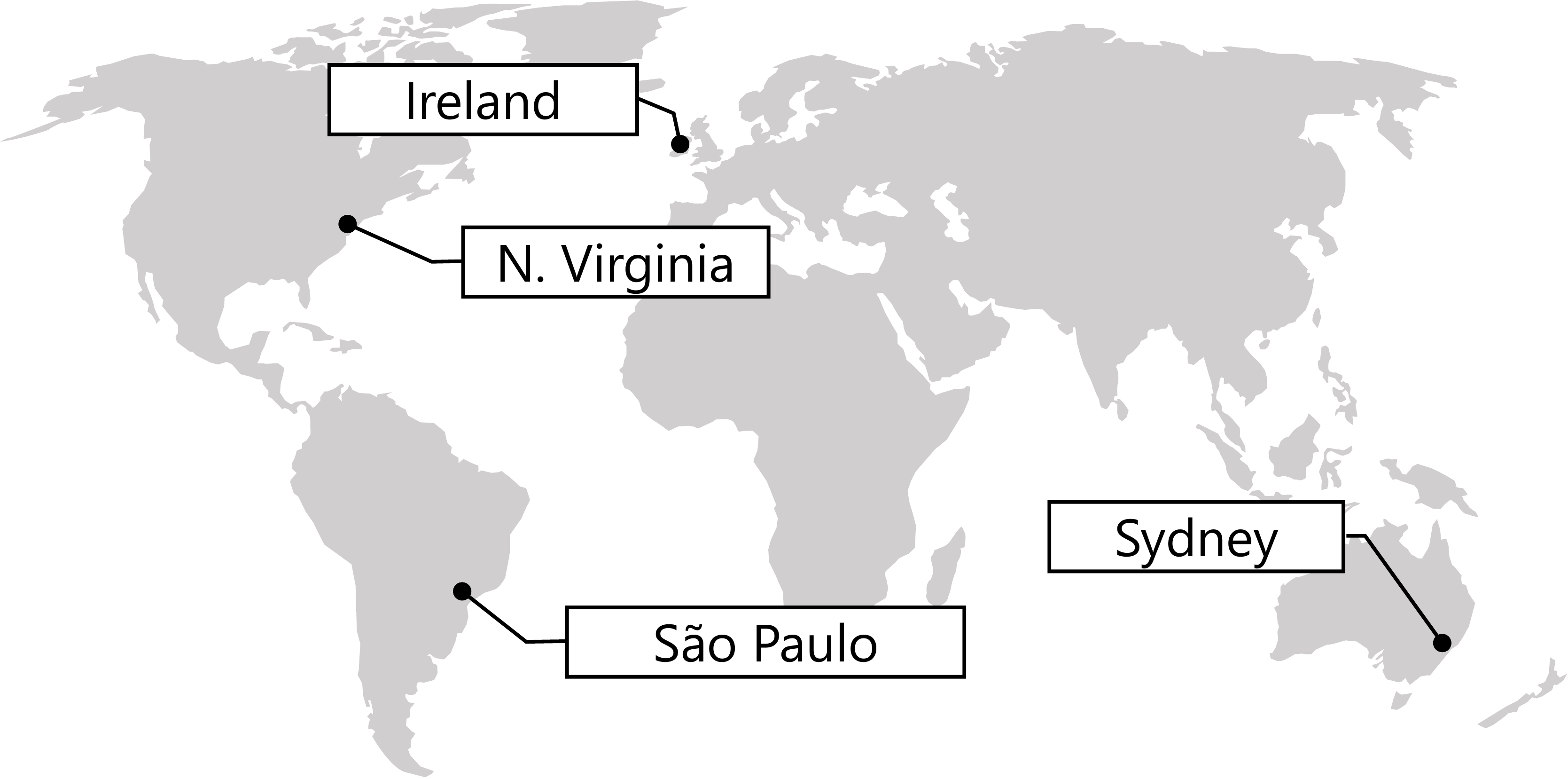}
        \label{fig:worldwide-group}
    }
    \subfloat[European Group]{
        \includegraphics[width=75mm]{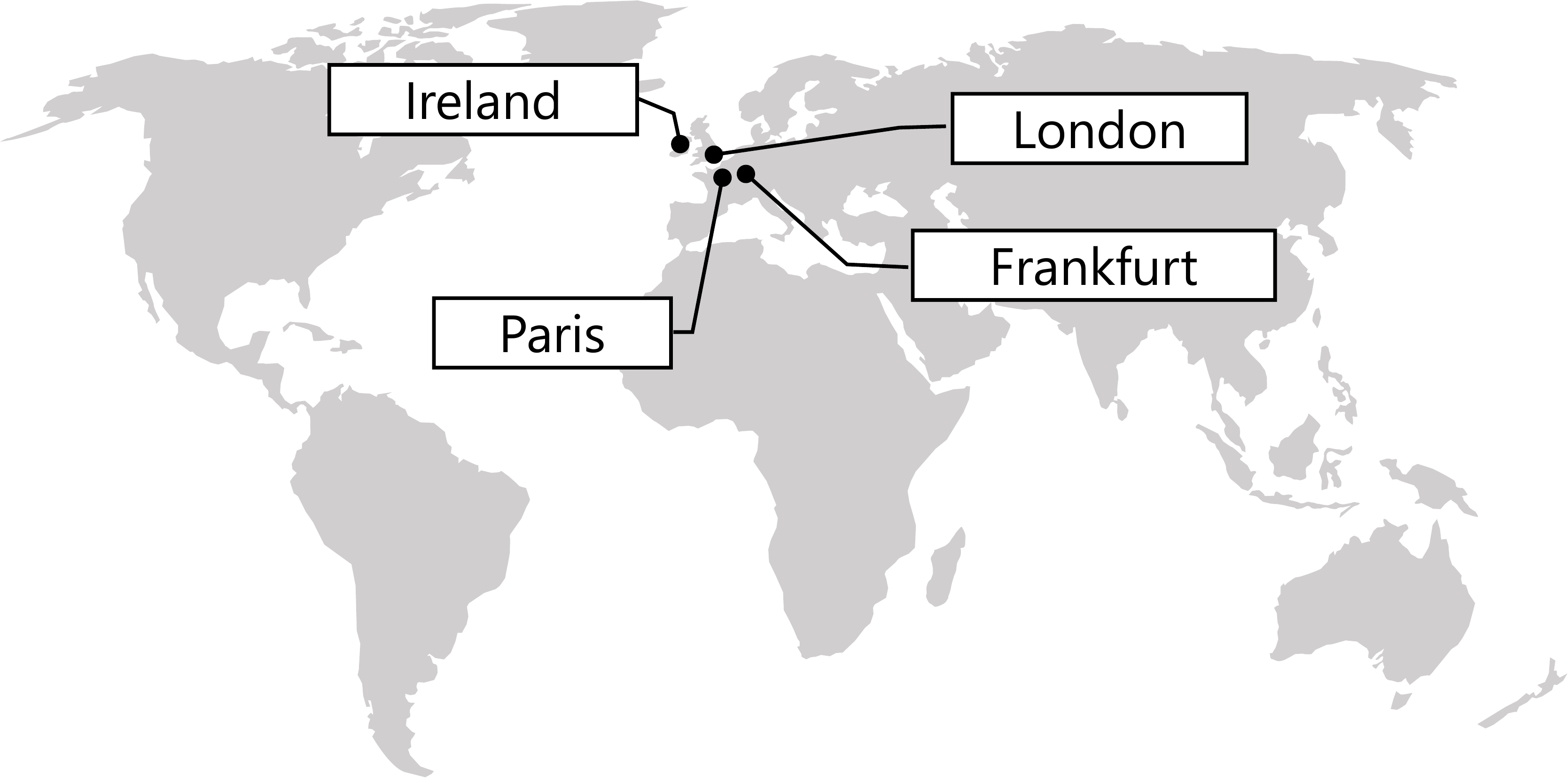}
        \label{fig:european-group}
    }
    \caption{Geographic location of replicas in each group.}
    \label{fig:group-map}
\end{figure}

In contrast, in geographic SMR, replicas communicate with each other over WAN, so the characteristics of the communication environment are different from those of conventional SMR.
Figure \ref{fig:bandwidth-change} shows the change in the communication bandwidth over seven days (Fig.~\ref{fig:bandwidth-change}\subref{fig:bandwidth-change-group-a-virginia} is from April 26 to May 3, 2021, and Fig.~\ref{fig:bandwidth-change}\subref{fig:bandwidth-change-group-b-london} is from June 1 to June 8, 2021) among Amazon EC2 regions commonly used in geographic SMR.
The communication bandwidth was measured hourly using \texttt{iperf} 3.1.3\footnote{\url{https://iperf.fr}}.
We deployed two groups of replicas, called Worldwide Group and European Group, into regions, as depicted in Fig.~\ref{fig:group-map}.
The Worldwide Group deploys replicas in four different continents: North Virginia, Ireland, São Paulo, and Sydney, while the European Group deploys replicas in European regions: London, Frankfurt, Ireland, and Paris.
In Worldwide Group (Fig.~\ref{fig:bandwidth-change}\subref{fig:bandwidth-change-group-a-virginia}), the communication bandwidth differs considerably among the replicas.
The degree of change in the communication bandwidth varies from region to region, with little changes in São Paulo and large changes in Ireland.
In contrast, in European Group(Fig.~\ref{fig:bandwidth-change}\subref{fig:bandwidth-change-group-b-london}), the difference in the average communication bandwidth among replicas is small but changes are very large for all regions.
Thus, in a geographic SMR, since the available communication bandwidth differs for each replica and changes frequently and largely, the existing state transfer methods cannot transfer the state efficiently.

In this paper, we propose a state transfer method that adapts to network bandwidth variations in geographic SMR.
The proposed method has the following features:

\textbf{Chunk division and bandwidth-based allocation:}
The proposed method divides the replicated service state into several small \emph{chunks} and changes the number of chunks allocated to a replica so that a replica with a broader communication bandwidth transfers more chunks than others.
Ideally, the state should be strictly divided based on communication bandwidths, but for this purpose, a recovery replica needs to know the current state size in advance.
However, since geographic SMR generally has high communication delays \cite{ThousandEyes2019}, increasing the number of communications will increase the state transfer time and complicate the state transfer process.
Therefore, the proposed method divides the whole state into small $N$ chunks, where $N$ is the predefined value, and a recovery replica requests, as a partial state, $m$ chunks so that $m$ is close to the ratio of the communication bandwidth between each replica.

\begin{figure}[t]
\centering
\includegraphics[width=75mm]{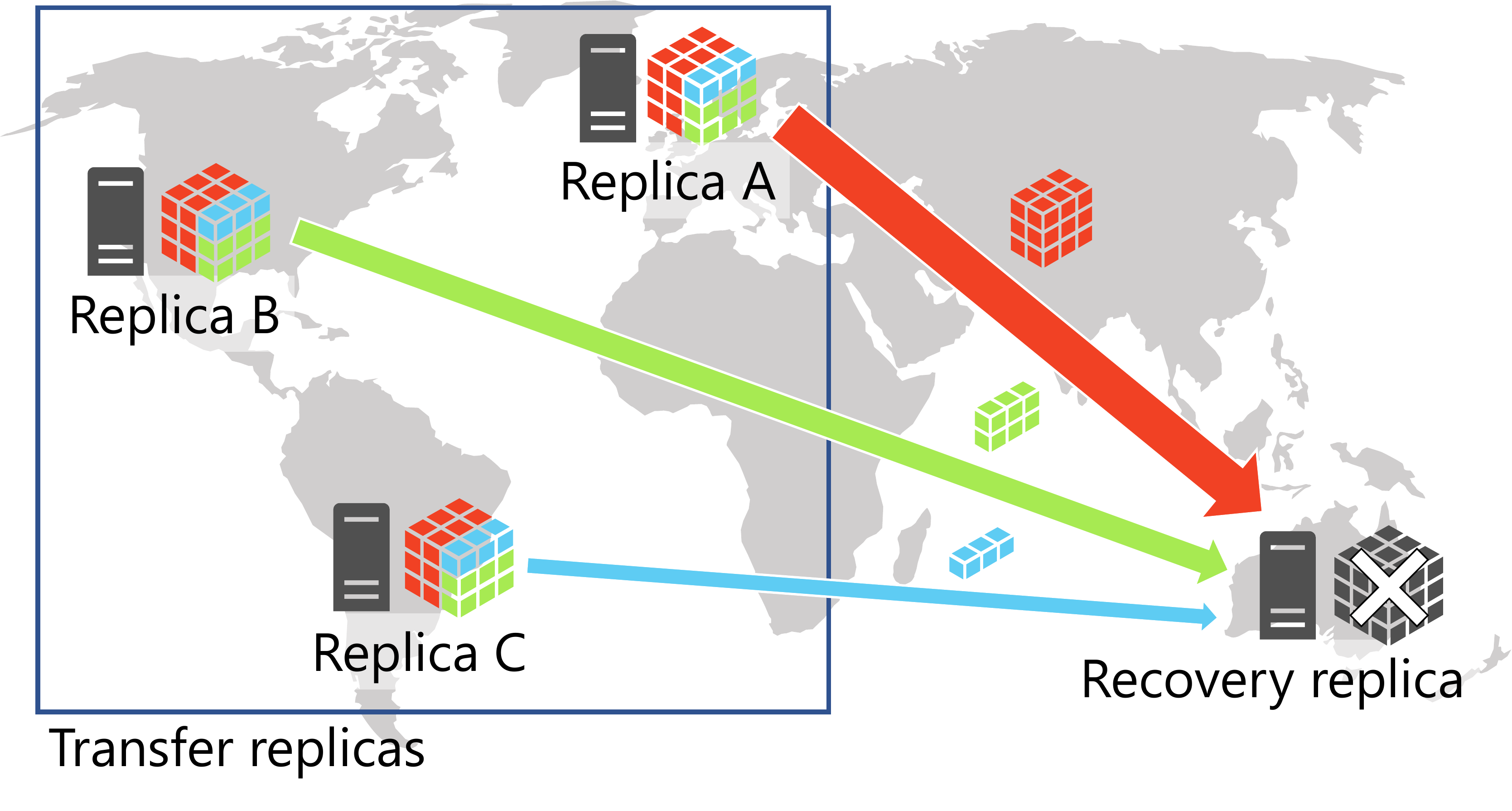}
\caption{An example of state transfer with the proposed method.}
\label{fig:proposed-fig}
\end{figure}

Figure \ref{fig:proposed-fig} shows an example of state transfer with the proposed method.
In Fig.~\ref{fig:proposed-fig}, three transfer replicas, A, B, and C, transfer chunks to the recovery replica.
Replicas A and C have the broadest and narrowest bandwidths to the recovery replica, respectively.
The recovery replica estimates the communication bandwidth with each replica and requests the number of chunks close to the ratio of their communication bandwidths (e.g., 18 chunks to replica A and 3 chunks to replica C).
The transfer replicas divide the state into chunks and transfer the chunks requested by the recovery replica.

\textbf{Updating allocated chunks according to changes in the communication bandwidth:}
In the proposed method, a recovery replica always measures the communication bandwidth with each replica based on the SMR message reception speed.
Since this measurement is passive, there is no overhead.
When the ratio of the communication bandwidth between replicas changes, the recovery replica periodically reallocates the chunks based on the new ratio to adapt to dynamic changes in the communication bandwidth.

\textbf{Per-chunk hashing and efficient integrity verification:}
When a recovery replica receives a service state, it needs to verify the correctness of the state to avoid applying a false state from a malicious replica.
The proposed method efficiently verifies the integrity of the received state by assigning and verifying the integrity per chunk.

For evaluation of the proposed method, we first show the superiority of the proposed method in geographic SMR analytically.
In the analysis, we use the variation in the bandwidth between each replica and the estimation error in bandwidth estimation of the proposed method as variables and formulate the state transfer times of the proposed and existing methods. 
Assuming that the average bandwidth between replicas is 100 Mbps, even if the standard deviation of the bandwidth is 20 Mbps or more and the estimation error is approximately 10\%, the proposed method is faster than existing methods.

We also show that the proposed method reduces the state transfer time of geographic SMR experimentally.
We build a geographic SMR system on Amazon EC2 and measure the state transfer time of the proposed method and existing methods.
The experimental results show that the proposed method can reduce the state transfer time by up to 47\% compared to the CST protocol in the environment with a large difference (Worldwide Group) and large changes (European Group) in the communication bandwidth.

Moreover, we examine the dynamic replacement of replicas using the proposed state transfer method. 
In geographic SMR, the latency between replicas varies depending on the geographic location of the replicas and, as a result, the latency of requests also varies \cite{Numakura2019b,Sousa2015}.
In addtion, the latency between the same replicas is not constant and varies over time \cite{Numakura2019b}.
These variations are caused by network congestion and changes in routing paths, which exacerbate the request latency.
\emph{Dynamic replica replacement} prevents exacerbation of the request latency due to latency changes between replicas by dynamically moving replicas during replication.
However, dynamic replica replacement in geographic SMR has not been attempted so far, because replica replacement requires state transfer processing, which is considered time-consuming in geographic SMR\cite{Sousa2015}.
In contrast, our proposed method can shorten the state transfer time in geographic SMR.
Because the proposed method is also expected to enhance the practicality of dynamic replica replacement, we examine this practicality using the proposed method.

The remainder of this paper is organized as follows.
Section \ref{sec:related-work} describes SMR and existing state transfer methods.
Section \ref{sec:proposed-method} proposes the state transfer method that adapts to network bandwidth variations in geographic SMR.
Section \ref{sec:analysis} theoretically analyzes the state transfer times of the proposed method and existing ones.
Section \ref{sec:evaluation} shows the performance evaluation results using a geographic SMR system built on Amazon EC2.
Section \ref{sec:replica-relocation} verifies the practicality of the dynamic replica replacement using the proposed method.
Section \ref{sec:conclusion} concludes the paper.

\section{Related work}
\label{sec:related-work}

\subsection{State Machine Replication (SMR)}
\label{sec:smr}

State Machine Replication (SMR) \cite{Schneider1990,Distler2021} is a method of replicating a service to multiple servers (replicas) and processing requests in the same order to keep the replicas in the same state; as a result, the fault tolerance of the service can be improved.
Although SMR requires that all replicas receive requests in the same order, the communication delay between the replicas is different in the actual network.
Even if a client sends a request to all replicas simultaneously, the request arrival time differs among the replicas.
In such a case, a distributed protocol, called Total Order Broadcast or Atomic Broadcast \cite{D'efago2004}, can ensure that each message is delivered to all participants in a group in the same order.
We can realize SMR by using Total Order Broadcast to deliver requests to replicas.

Most SMR protocols \cite{Bessani2014,Castro1999,Castro2002,Junqueira2011,Kotla2010,Lamport1998,Ongaro2014, Sousa2012} elect one leader among the replicas, which determines the processing order of requests and distributes the decision to other replicas.
Even if the leader fails, the replicas must not process the requests in a different order.
To meet this requirement, the replicas agree with the processing order \cite{Lamport1982}.
The difficulty of agreement with faulty replicas varies depending on an assumed failure model, and the number of necessary replicas for agreement also changes.
Hereafter, we denote the total number of replicas and the upper bound of faulty replicas by $n$ and $f$, respectively.

According to their target failure model, we can classify SMR protocols into Crash Fault-Tolerant SMR (CFT-SMR) and Byzantine Fault-Tolerant SMR (BFT-SMR).
CFT-SMR resists crash faults in which a faulty replica does not operate at all after the failure, while BFT-SMR resists Byzantine fault \cite{Lamport1982}, in which a faulty replica behaves arbitrarily without following the protocol after the failure.
There are several protocols for CFT-SMR, such as Paxos \cite{Lamport1998}, Raft \cite{Ongaro2014}, and ZAB \cite{Junqueira2011}, as well as for BFT-SMR, such as PBFT \cite{Castro1999}, Zyzzyva \cite{Kotla2010}, and BFT-SMaRt \cite{Sousa2012}.
CFT-SMR and BFT-SMR protocols require $n \geq 2f+1$ \cite{Lamport1998,Chandra1996} and $n \geq 3f+1$ \cite{Lamport1982}, respectively.

\subsection{Geographic SMR}

Geographic SMR places replicas at a large distance from each other and tolerates large-scale disasters, such as earthquakes.
Since geographic SMR exhibits different characteristics from conventional SMR, there are various SMR protocols specialized for geographic SMR \cite{Amir2010,Berger2019,Coelho2018,Coelho2021,Eischer2018,Eischer2020,Eischer2021,Enes2020,Enes2021,Liu2017,Mao2008,Nawab2018,Ngo2020,Sousa2015,Xu2019,Yan2020,Zhao2018}.
The most significant difference between SMR in a data center and geographic SMR is the inequality of latency and throughput between replicas and clients.
In geographic SMR, latency (communication delay) \cite{Numakura2019b,Sousa2015} and throughput (Fig.~\ref{fig:bandwidth-change}) between replicas differ greatly from replica to replica.
However, since the SMR protocols designed for a data center assume that messages arrive at almost the same time on all replicas, these SMR protocols cannot demonstrate their original performance in geographic SMR.
Therefore, many SMR protocols designed for geographic SMR \cite{Mao2008,Zhao2018,Sousa2015,Berger2019,Liu2017,Eischer2018,Coelho2018,Ngo2020,Coelho2021,Yan2020,Eischer2021,Enes2020,Enes2021} assume different latencies and use various techniques to overcome this difference.

WHEAT \cite{Sousa2015} solves this problem and shortens the time required for consensus by adding extra replicas to the system and changing the voting weight in the consensus process for each replica.
AWARE \cite{Berger2019} responds to changes in the network environment by dynamically changing the voting weight and location of the leader in the consensus.
Steward \cite{Amir2010}, Weave \cite{Eischer2020}, and Geo-Raft \cite{Xu2019} build SMRs hierarchically in two layers, inside each data center and between the data centers.
This reduces communication over a slow WAN, and, as a result, these protocols can reduce latency.
As explained in Sect.~\ref{sec:smr}, many SMR protocols require a leader to order the requests, in which replicas must forward the requests to the leader.
Since this transfer adversely affects latency, geographic SMR often uses leaderless protocols \cite{Eischer2021,Enes2020,Enes2021}.
In a leaderless protocol, any replica can determine the processing order of requests by coordinating with other replicas.
The features and limitations of leaderless SMR protocols were investigated by Rezende et al.~\cite{Rezende2020}.
In geographic SMR, the processing performance of replicas (e.g., CPU and memory) might not be uniform.
Even if they are the same, they might not be able to achieve their original performance due to network failures or load congestion.
In this case, the slow replica becomes the bottleneck, and the service latency increases.
To avoid this, there are protocols that distribute SMR functions to replicas to balance loads \cite{Zhao2018,Ngo2020,Xu2019}.
There is also a method to find the best replica deployment based on the latency between replicas and clients \cite{Numakura2019b}.

The proposed state transfer method targets geographic SMR and can be used in combination with various geographic SMR protocols, such as WHEAT and AWARE.
This method supports both CFT-SMR and BFT-SMR.

In Sect.~\ref{sec:replica-relocation}, we apply the proposed method to dynamic replica replacement, which improves service latency by moving a replica to another location.
This replacement can be used complementarily with geographic SMR protocols in situations where it is difficult to improve latency with an adaptive function, such as AWARE \cite{Berger2019} or where we want to place replicas based on the suggestion obtained by the method by Numakura et al.~\cite{Numakura2019b} during replication.

\subsection{State Transfer of SMR}
\label{sec:state-transfer}

In SMR, to restore the failed replica to replication as a normal replica again, the replica transfers the latest state to the failed replica \cite{Schneider1990}.
This cooperation is called \emph{state transfer} and we explain two major state transfer methods here.

\begin{figure}[t]
    \centering
    \includegraphics[width=75mm]{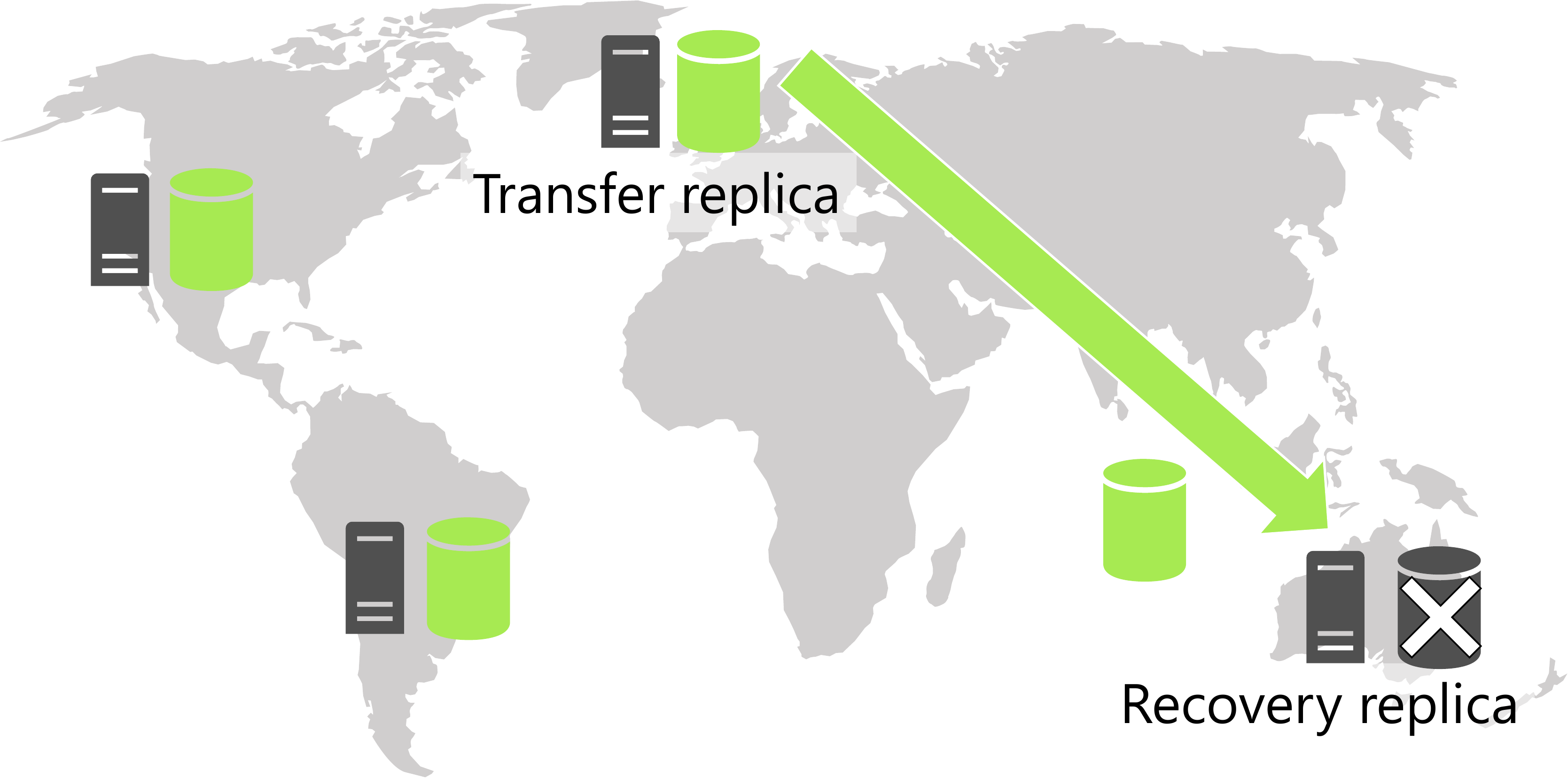} 
    \caption{State transfer of PBFT protcool.}
    \label{fig:chunk-transfer-schneider}
\end{figure}

The most basic state transfer method is one in which a single replica transfers the entire state to the failed replica.
This method was introduced by Schneider \cite{Schneider1990} and adopted by various SMR protocols, such as PBFT \cite{Castro1999,Castro2002}.
Hereafter, we call this method \emph{PBFT state transfer}.
An example of PBFT state transfer is shown in Fig.~\ref{fig:chunk-transfer-schneider}.
In PBFT state transfer, a state consist of two types of data: \emph{checkpoint} and \emph{log}.
The checkpoint is a replica of the current service status and acquired periodically, and the log is a list of requests executed after the latest checkpoint is acquired.
The checkpoint and log are transferred simultaneously; thus, in the explanation of PBFT state transfer, we just call them a state.

We now explain the flow of PBFT state transfer.
First, the recovery replica sends a request to transfer the latest state to one of the replicas, called transfer replica, and the transfer replica transfers its entire state to the recovery replica.
In the case of BFT-SMR, the transfer replica might be Byzantine fault and transfer an incorrect state.
To prevent this, the recovery replica requests the hash value of the entire state from replicas other than the transfer replica.
The recovery replica uses these hash values to verify the integrity of the received state as follows.
First, the recovery replica compares the received hashes with the hash $h$ calculated from the state received from the transfer replica.
If the recovery replica receives the hashes that match $h$ from $f + 1$ or more replicas, the recovery replica decides that the replicas include at least one non-faulty replica and concludes that the received state is correct.

When the recovery replica receives the state from the transfer replica and the hash verification is successful (in the case of BFT-SMR), the recovery replica applies the state to itself and completes the state transfer.
If the transfer replica does not transfer a state or the hash verification fails (in the case of BFT-SMR), the recovery replica again requests the status transfer from a replica other than the current transfer replica.

PBFT state transfer is performed by only two replicas (one recovery replica and one transfer replica).
Therefore, when a client sends a request to a service on SMR during a state transfer, the replicas that are not involved in the state transfer maintain their normal processing performance.
In contrast, the recovery replicas and transfer replica must devote their processing power to both state transfer and request processing.
Therefore, the processing performance of the service on SMR decreases during the status transfer \cite{Bessani2013}.

\begin{figure}[t]
    \centering
    \includegraphics[width=75mm]{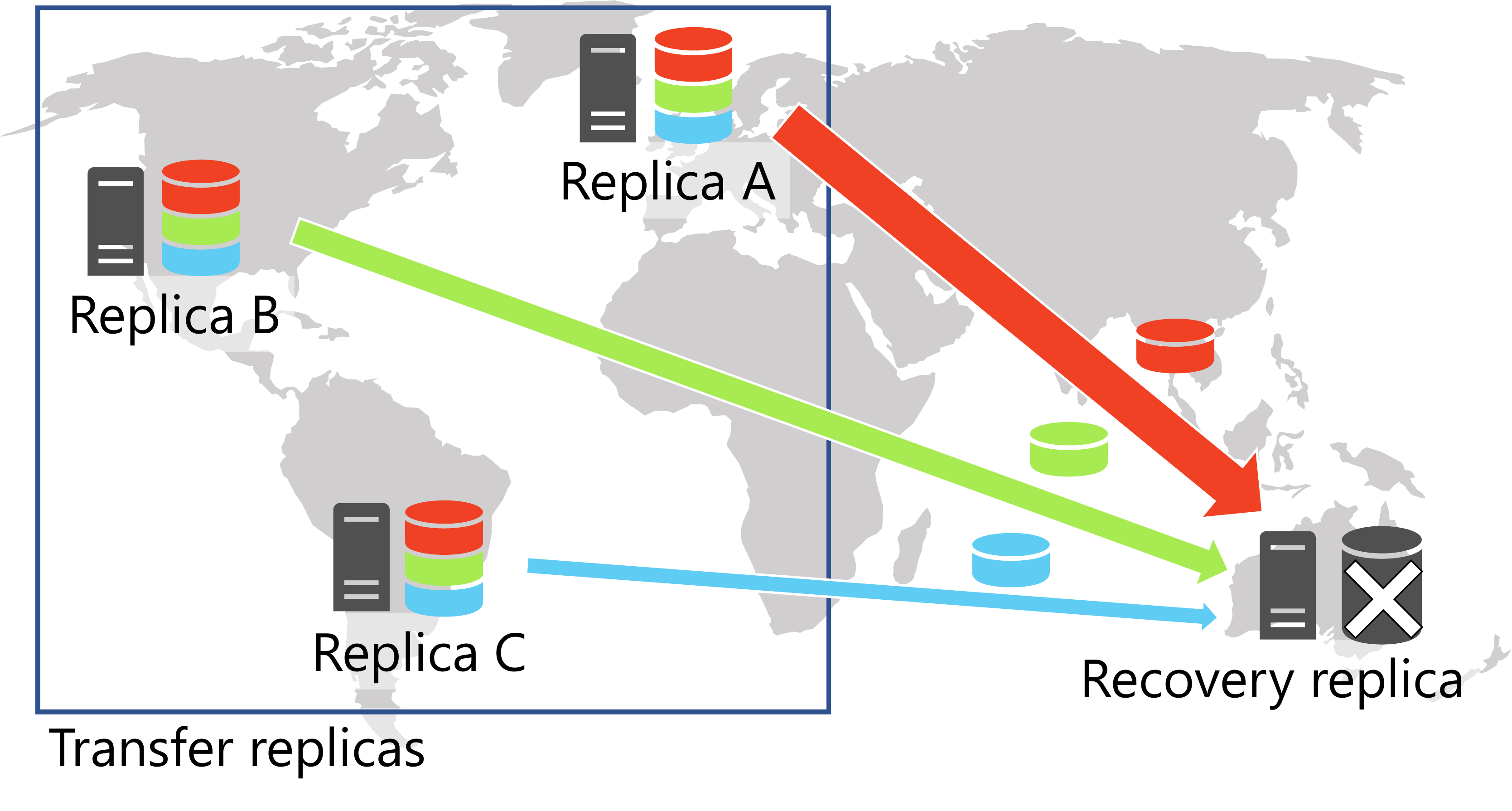} 
    \caption{State transfer of CST protocol.}
    \label{fig:chunk-transfer-bessani}
\end{figure}

In order to alleviate the degrading processing performance of a service during a state transfer, Bessani et al.~proposed a fast state transfer method, called Collaborative State Transfer (CST) protocol \cite{Bessani2013}.
This method shortens the transfer time by dividing a state into multiple sub-states (checkpoints and logs) and transferring them from multiple replicas cooperatively.
Figure \ref{fig:chunk-transfer-bessani} shows an example of state transfer using the CST protocol.

In the CST protocol, the recovery replica requires state transfer to all other replicas with the latest state (replicas A, B, and C in Fig.~\ref{fig:chunk-transfer-bessani}).
When a state is requested, one transfer replica transfers the entire checkpoint, while the other transfer replicas send an equally-divided part of a log simultaneously.
Figure \ref{fig:chunk-transfer-bessani} assumes the case in which the checkpoint and equally divided log have the same size.
In the case of BFT-SMR, to verify the integrity of the received state, the transfer replicas also send the hashes of the checkpoint and the divided log.
If the recovery replica receives a checkpoint and all divided logs and succeeds in the hash verification (in the case of BFT-SMR), it applies the state to itself and completes the state transfer.

\section{Proposed Method}
\label{sec:proposed-method}

Here, we describe the proposed state transfer method.
We present the overview of this method in Sect.~\ref{sec:overview} and the details in Sect.~\ref{sec:details}.
Finally, we prove the correctness of this method in Sect.~\ref{sec:correctness}.

\subsection{Overview}
\label{sec:overview}

\begin{figure}[t]
    \centering
    \includegraphics[width=70mm]{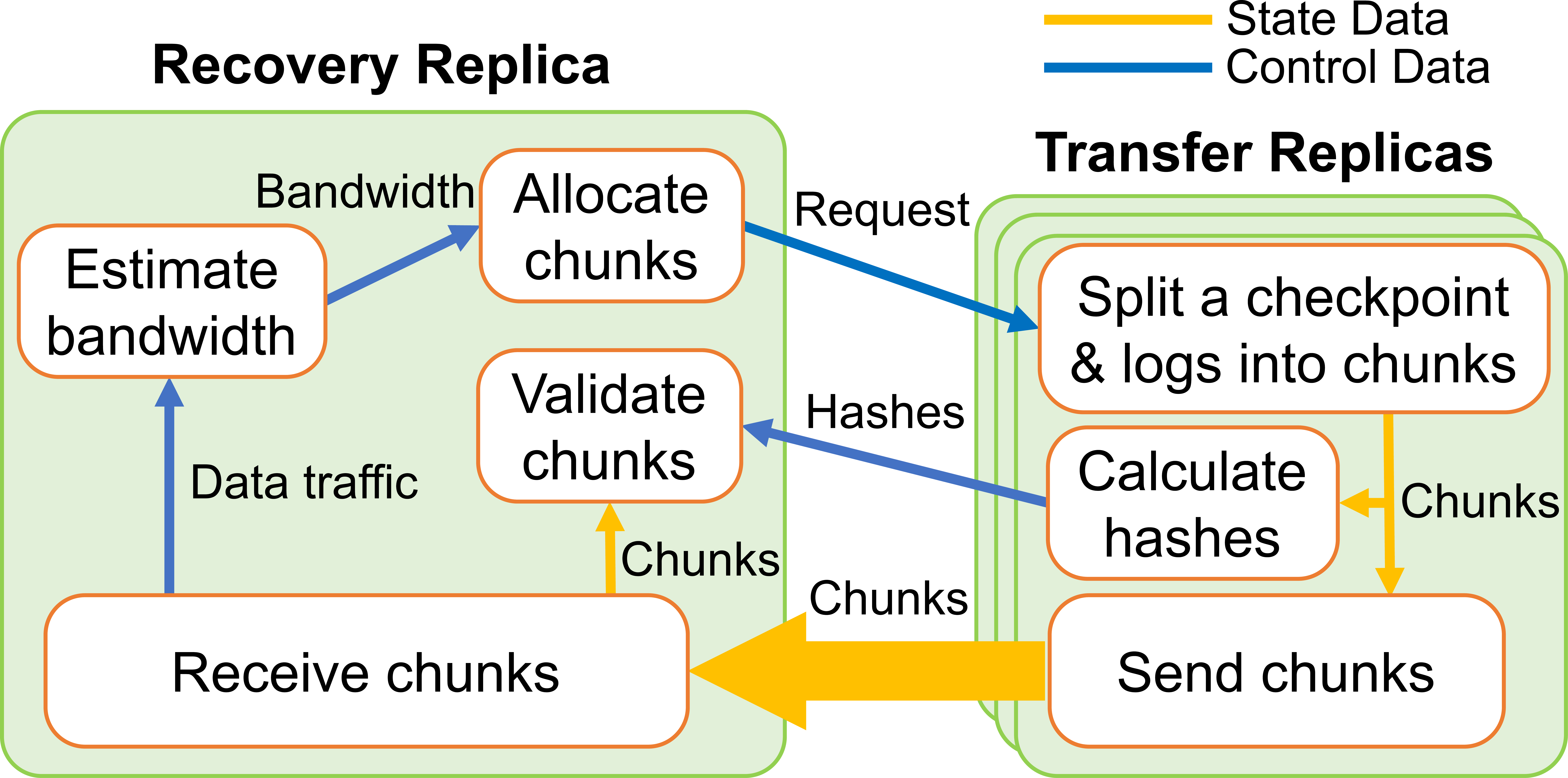} 
    \caption{Overview of the proposed method.}
    \label{fig:proposed-block}
\end{figure}

Figure \ref{fig:proposed-block} shows the overview of the proposed method.
The proposed method internally represents the service state by the checkpoint and the log as in the CST protocol, but during a state transfer, the state is represented as the binary data obtained by combining them.
Furthermore, the state is divided into $N$ chunks of the same size and transferred in these units.
In this way, the amounts of data transferred by each replica can be easily adjusted.

A recovery replica assigns $N$ chunks to the transfer replicas at the start of the state transfer and requests the transfer of the chunks from them.
When a transfer replica receives the chunk transfer request, it divides the checkpoint and the log into chunks.
The transfer replica then sends the chunks and their hashes, which can be used to verify the integrity of the chunks, to the recovery replica.

When the recovery replica receives a chunk from the transfer replica, it calculates a hash of the received chunk and compares the hash it calculated with the hash it received from the transfer replica to exclude the fake chunk sent by a Byzantine replica.
This verification proceeds in parallel with the sending and receiving processes for the state transfer.
This reduces the state transfer time and enables early detection of fake chunks and their re-request.

The recovery replica always records the reception rate of chunks from each transfer replica during a state transfer and estimates the communication bandwidth based on the rate.
Based on the estimated bandwidth, the recovery replica adjusts the chunk allocation of the transfer replicas such that the replica with a broader communication bandwidth than others transfers more chunks.
In this way, the proposed method adapts to dynamic changes in the communication bandwidth.

When the recovery replica receives all chunks and completes the verification, it restores the checkpoint and log from the chunks and applies them to the replica along with the received log during the state transfer to restore the latest service state and finishes the state transfer.

\subsection{Details}
\label{sec:details}

Here, we describe the details of the proposed method.
First, we introduce some notations to describe the method formally.
In the proposed method, all replicas except the recovery replica $r$ behave as transfer replicas.
We denote the set of all transfer replicas by $T$.
The recovery replica repeatedly requests a set of chunks from the transfer replica every $I$ seconds.
The $i$-th chunk set that the recovery replica $r$ requests from the transfer replica $t$ is denoted by $C_t^i$, and the set of all chunks that represents the latest service state is denoted by $C_{all}$.
The average communication bandwidth with the transfer replica $t$ estimated by the recovery replica between the ($i-1$)-th and $i$-th times is denoted as $w_t^i$, and the sum of the communication bandwidth with all transfer replicas is denoted as $w_{all}^i=\sum_{j \in T}w_j^i$.
In the case of $i=0$, a recovery replica cannot estimate the bandwidth; thus, we assume $w_t^0=1$ for every transfer replica $t$.

In BFT-SMR, Byzantine replicas might transfer fake chunks to a recovery replica as transfer replicas.
To avoid such attacks, the recovery replica collects hashes of each chunk and verifies the integrity of the received chunks.\footnote{In CFT-SMR, a faulty replica just stops processing. Therefore, a recovery replica does not need to take care of such attacks.}
The set of hashes of all chunks is denoted as $H_{all}$, the set of hashes for which a recovery replica verifies the integrity of their chunks is denoted as $H$, and the hash function is denoted as $\mathit{Hash}(x)$.

Algorithm \ref{alg:recovery-replica} shows the pseudocode of a recovery replica for BFT-SMR.
When a state transfer is required, a recovery replica $r$ starts two tasks: $T_1$ to request chunks and $T_2$ to receive chunks.
First, task $T_1$ requests hashes $H_{all}$ of all chunks from all transfer replicas $T$ (lines \ref{algl:t1_hash_req_begin}--\ref{algl:t1_hash_req_end}).
Next, $r$ requests a set of chunks $C_{all}$ from $T$ every $I$ seconds until all chunks $C_t^i$ are received (lines \ref{algl:t1_chunk_req_begin}--\ref{algl:t1_chunk_req_end}).
Here, $r$ determines $C_t^i$ such that it satisfies the conditions
\begin{equation}
    \label{eq:chunk_cond_1}
    |C_t^i| = (N - |C|) \cdot w_t^i/w_{all}^i
\end{equation}
\begin{equation}
    \label{eq:chunk_cond_2}
    C \cup \bigcup_{j \in T} C_j^i = C_{all}, 
\end{equation}
where $C$ is the set of chunks that $r$ has received and verified.
If $|C_t^i|=0$ because of a narrow bandwidth, a recovery replica allocates one chunk, which overlaps with $C_a^i$ of another transfer replica $a$.\footnote{This is because some data transfer is necessary to passively estimate the communication bandwidth.}

Task $T_2$ continues to receive chunks from the transfer replicas until it receives all chunks $C_{all}$ (lines \ref{algl:t2_chunk_req_begin}--\ref{algl:t2_chunk_req_end}).
After receiving a chunk $c$ that has not been verified, a recovery replica $r$ verifies its integrity (lines \ref{algl:t2_chunk_verify_begin}--\ref{algl:t2_chunk_verify_end}).
The verification succeeds if $r$ receives the same hash from $f+1$ transfer replicas and $h$ matches hash $h' = \mathit{Hash}(c)$, which is locally computed in $r$ from $c$.
If the verification succeeds, $r$ includes $c$ in $C$.
When the verification fails, the recovery replica $r$ requests $c$ again from another transfer replica different from the previous sender of $c$ in the next chunk request.
If the recovery replica $r$ receives hashes from $|T|-f$ transfer replicas, but at least $f+1$ replicas of them do not send the same hash, this method switches to the PBFT state transfer method.
After receiving all the chunks, $T_2$ combines $C$ to obtain the checkpoint and log and uses them to restore the latest state (lines \ref{algl:t2_state_recovery_begin}--\ref{algl:t2_state_recovery_end}).
When the recovered replica is up-to-date, the proposed method terminates.

Algorithm \ref{alg:transfer-replica} shows the pseudocode of a transfer replica for BFT-SMR.
When a transfer replica $t$ receives a request for hashes of all chunks from a recovery replica $r$, $t$ calculates $H$ from the latest checkpoint and log and sends it to $r$ (lines \ref{algl:hash_transfer_begin}--\ref{algl:hash_transfer_end}).
When a received request wants a chunk set $C_t^i$, $t$ divides the checkpoint and log into chunks and transfers the requested chunks to $r$ one by one (lines \ref{algl:chunk_transfer_begin}--\ref{algl:chunk_transfer_end}).

In the case of CFT-SMR, the following processes are changed.
First, in Task $T_1$, the request of $H$ (lines \ref{algl:t1_hash_req_begin}--\ref{algl:t1_hash_req_end} of Alg.~\ref{alg:recovery-replica}) is no longer required.
Next, Task $T_2$ skips the integrity verification of the received chunk $c$ (lines \ref{algl:t2_chunk_verify_begin}--\ref{algl:t2_chunk_verify_end} of Alg.~\ref{alg:recovery-replica}) and simply adds $c$ to $C$.
Finally, a transfer replica is never requested to transfer $H$ (lines \ref{algl:hash_transfer_begin}--\ref{algl:hash_transfer_end} of Alg.~\ref{alg:transfer-replica}).

\begin{algorithm}[t]

\caption{The pseudocode of the recovery replica}
\label{alg:recovery-replica}

\begin{algorithmic}[1]
\small
\Statex \hspace{-2em} Initialization:
\State $C \gets \emptyset$
\State \textbf{Activate task} $T_1$, $T_2$
\vspace{2mm}

\Statex \hspace{-2em} Task $T_1$:
\State $i \gets 0$
\For{$t$ \textbf{in} $T$} \label{algl:t1_hash_req_begin}
    \State Request the set $H_{all}$ of all hashes to $t$
\EndFor \label{algl:t1_hash_req_end}
\While{$C \neq C_{all}$} \label{algl:t1_chunk_req_begin}
    \For{$t$ \textbf{in} $T$}
    \State Calculate $C_{t}^i$ from Equations.~\eqref{eq:chunk_cond_1} and \eqref{eq:chunk_cond_2}
        \State Request $C_{t}^i$ to $t$
    \EndFor
    \State Wait $I$ seconds
    \State $i \gets i + 1$
\EndWhile \label{algl:t1_chunk_req_end}
\vspace{2mm}

\Statex \hspace{-2em} Task $T_2$:
\When{receive a set of hashes $H_{all}$ from the transfer replica $t \in T$}
    \If{it has received $H_{all}$ from $|T|-f$ or more transfer replicas and it has not received more than $f+1$ same $H_{all}$}
        \State Switch to the state transfer method of PBFT protocol
    \EndIf
\EndWhen
\When{a chunk set $C_t \subset C_{all}$ is delivered from a recovery replica $t \in T$}
    \For{$c$ \textbf{in} $C_t \setminus C_{all}$} \label{algl:t2_chunk_req_begin}
        \If{it has already received the same $f+1$ hash $h$ of chunk $c$ and $h$ matches $h' = \mathit{Hash}(c)$}
        \label{algl:t2_chunk_verify_begin}
            \State $C \gets C \cup \{c\}$
        \EndIf \label{algl:t2_chunk_verify_end}
    \EndFor \label{algl:t2_chunk_req_end}
    \If{$C = C_{all}$}
        \State Combine $C$ and get checkpoint $a$ and log $L$ \label{algl:t2_state_recovery_begin}
        \State Apply checkpoint $a$
        \State Process log $L$ and the log received during state transfer in order of oldest to newest \label{algl:t2_state_recovery_end}
        \State \textbf{Terminate}
    \EndIf
\EndWhen

\end{algorithmic}
\end{algorithm}

\begin{algorithm}[t]

\caption{The pseudocode of the transfer replica}
\label{alg:transfer-replica}
\begin{algorithmic}[1]
\small

\When{the hash set $H_{all}$ of all chunks is requested from a recovery replica $r$} \label{algl:hash_transfer_begin}
    \State Transfer $H_{all}$ to $r$
\EndWhen \label{algl:hash_transfer_end}
\vspace{2mm}

\When{a chunk set $C_t^i$ is requested from a recovery replica $r$} \label{algl:chunk_transfer_begin}
    \State $C_t \gets C_t^i$
    \For{$c$ \textbf{in} $C_t$}
        \State Transfer $c$ to $r$
    \EndFor
\EndWhen \label{algl:chunk_transfer_end}

\end{algorithmic}
\end{algorithm}

\subsection{Correctness}
\label{sec:correctness}

We prove the correctness of the proposed method in terms of the following requirements.
\begin{description}
    \item[Safety:] The state applied to a recovery replica is the same as that of a non-faulty replica.
    \item[Liveness:] State transfer eventually terminates.
\end{description}

For safety, a recovery replica verifies the integrity of a chunk $c$ using identical hashes $h$ received from $f+1$ transfer replicas.
In addition, the recovery replica locally computes $h' = \mathit{Hash}(c)$ and compares it to $h$ to deal with a Byzantine transfer replica that might send a fake chunk and the correct hash.
This ensures that the chunk is identical to that of at least one non-faulty transfer replica. 
Therefore, the state applied to the recovery replica is the same as that of a non-faulty replica.

Next, we prove the liveness.
A recovery replica $r$ needs to receive all hashes and chunks and verify their integrity before the state transfer terminates.
For receiving hashes, even if $f$ transfer replicas are faulty, there are still at least $f+1$ non-faulty transfer replicas.
If the non-faulty replicas send a hash to the recovery replica when they process the same request, the recovery replica can receive the same hash from at least $f+1$ replicas.
However, the recovery replica may not be able to receive the same hash from $f+1$ replicas because the arrival timing of the status transfer request is different for each replica.
In such a case, the recovery replica switches to the PBFT status transfer method to ensure the liveness.
For receiving chunks, a Byzantine transfer replica may not send any chunk.
However, a recovery replica $r$ continues to request at least one chunk from each transfer replica.
Thus, even if the Byzantine transfer replica does not send any chunk, $r$ eventually receives all chunks from non-faulty transfer replicas.
Therefore, the state transfer eventually terminates.

\section{Analysis of State Transfer Time}
\label{sec:analysis}
In this section, we analytically compare the state transfer times of the proposed and existing methods under worst-case conditions.

\subsection{Formulation of State Transfer Time}

We consider a situation of state transfer to the recovery replica $r$, and let $S$ be the state size of a transfer replica $t$ at that time.
We denote the communication bandwidth from a transfer replica $t$ to the recovery replica $r$ as $w_{t,r}$.

First, we formulate the state transfer time of PBFT.
In PBFT state transfer, assuming that the recovery replica $r$ chooses the replica $t$ with the widest communication bandwidth as the transfer replica, the state transfer time is expressed as
	\begin{equation}
    \label{eq:pbft}
    T_\mathrm{PBFT} = \frac{S}{w_{t,r}}.
\end{equation}

In the CST protocol, we assume that all transfer replicas send partial states of the same size to the recovery replica $r$ in parallel.
This is the best case for the CST protocol, and the state transfer time can be expressed as 
	\begin{equation}
    \label{eq:cst-protocol}
    T_\mathrm{CST} = \frac{S}{(n-1) \cdot w},
\end{equation}
where $w=\min(w_{t_1,r}, w_{t_2,r}, \dots, w_{t_{n-1},r})$.

Next, let us consider the state transfer time of the proposed method.
If we assume that a recovery replica can measure the communication bandwidth of each transfer replica without any error, the state transfer time of the proposed method can be expressed as
\begin{equation}
    \label{eq:proposed-without-error}
    T_\mathrm{proposed} = \frac{S}{w_{all}},
\end{equation}
where $w_{all}=\sum_{i=1}^{n-1} w_{t_i,r}$.

However, in reality, it is difficult to measure the communication bandwidth without any error.
If there are errors in the estimated communication bandwidth, a replica with a narrow communication bandwidth might have to transfer more chunks, which increases the total transfer time longer.
To take such error into account, we assume the worst case, where all replicas are mistakenly overestimated (or underestimated) their communication bandwidth by $x$ \% compared to the actual bandwidth.
In this case, if the communication bandwidth faster than the average is underestimated by $x$ \% and that slower than the average is overestimated by $x$ \%, the state transfer time becomes the longest and can be expressed as
	\begin{equation}
    \label{eq:proposed-with-error}
    T_\mathrm{proposed\_w\_err} = \frac{S \cdot (1+x/100)}{w_{err\_all}},
\end{equation}
where $w_{err\_all}$ is the sum of the communication bandwidth between each transfer replica and the recovered replica $r$, including the estimation error.

\subsection{State Transfer Times in Different Communication Bandwidths}

\begin{figure}[t]
    \centering
    \includegraphics[width=70mm]{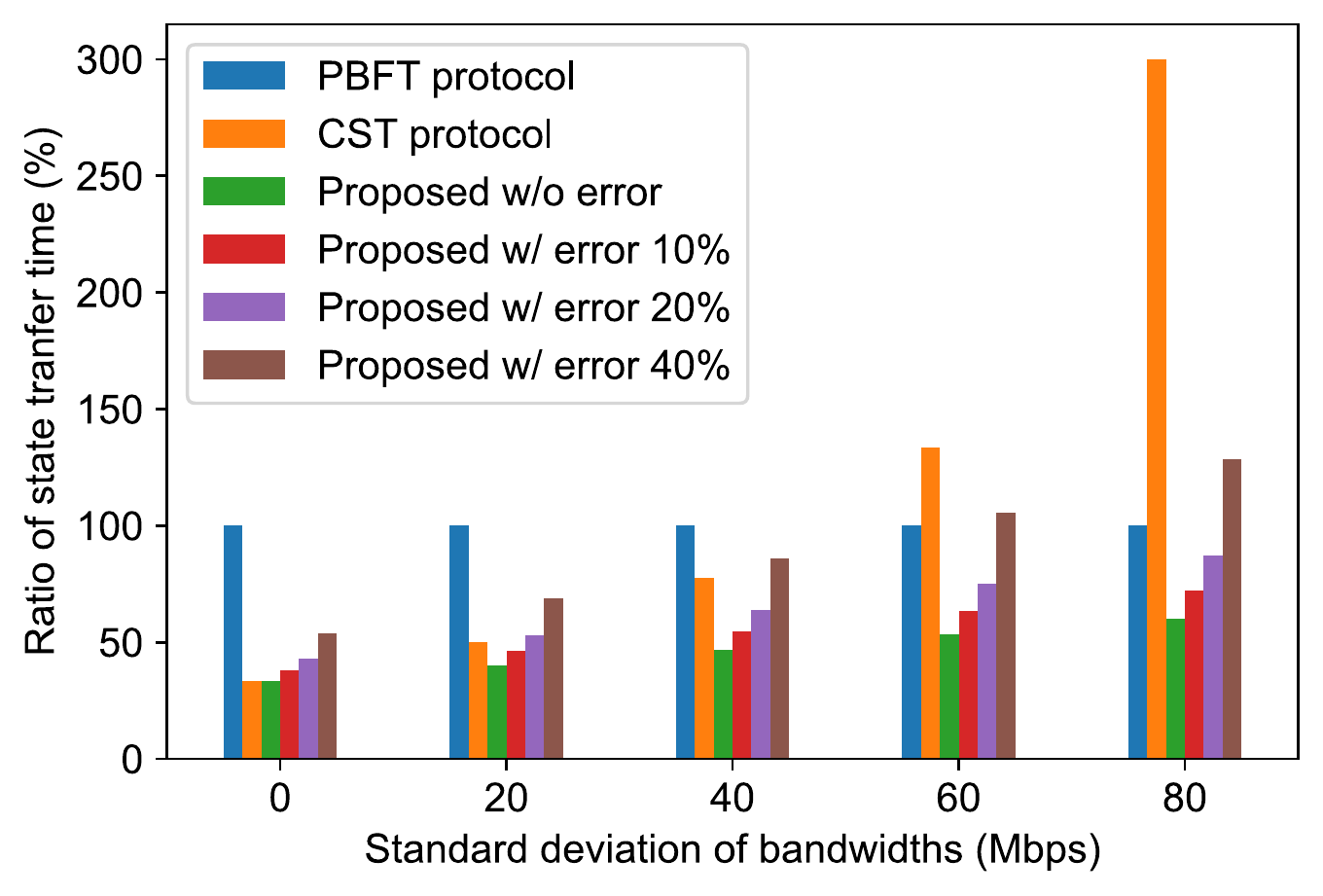} 
    \caption{State transfer times in different communication bandwidths and errors.}
    \label{fig:state-transfer-analysis}
\end{figure}

Figure \ref{fig:state-transfer-analysis} shows the state transfer times for each method in different communication bandwidths calculated by Equations \eqref{eq:pbft}--\eqref{eq:proposed-with-error}.
The figure assumes a situation in which the number of replicas is $n=4$ and there are no faulty replicas.
The average communication bandwidth between replicas is set as 100 Mbps, and the standard deviation ranges from 0 to 80 Mbps.
For example, if the standard deviation is 40 Mbps, the communication bandwidths of the three transfer replicas are 60, 100, and 140 Mbps, respectively.
The state transfer time is normalized to the PBFT transfer time as 100\%.

As shwon in Fig.~\ref{fig:state-transfer-analysis}, when the standard deviation increases, the state transfer time of the CST protocol increases compared to that of the PBFT protocol.
In the PBFT protocol, the replica with the widest communication bandwidth transfers the state; thus, PBFT can transfer the state at high speeds when the standard deviation of the communication bandwidth is large.
In contrast, each transfer replica sends the state equally in the CST protocol.
In other words, each transfer replica must send the same amount of data regardless of its communication bandwidth.
Therefore, in environments with large variations in the communication bandwidth, the transfer replica with a narrow bandwidth needs more time to send its partial state, and the total state transfer time also increases.
This tendency is clearly evident when the standard deviation of the communication bandwidth is 80 Mbps; in this case, the state transfer time is three times longer than that of PBFT.

The proposed method has the shortest state transfer time in the absence of any estimation error, and increases in the standard deviation do not affect the transfer time.
The proposed method allocates the amount of data to be transferred (i.e., the number of chunks) according to the communication bandwidth of each transfer replica so that the transfer time of each transfer replica is the same.
Therefore, if the total communication bandwidth of each transfer replica does not change, the transfer time will remain constant.
However, if there are estimation errors, the transfer time of each transfer replica does not align and deteriorates.
For example, when the estimation error is 40\%, the state transfer time is 1.6 times longer than that in the caes where the standard deviation of the communication bandwidth is 0 Mbps and 2.1 times longer when the standard deviation is 80 Mbps, compared to the case without any estimation error.
Although the transfer time of the proposed method is still shorter than that of the other methods in many cases, the proposed method becomes slower than PBFT when the standard deviation of the communication bandwidth is greater than 60 Mbps and the estimation error is 40\%.
However, such large estimation errors are rare, as we see in Sect.~\ref{sec:bandwidth-estimation-error}.

From these results, we have analytically confirmed that the proposed method can perform a state transfer at high speed in environments with large differences in communication bandwidth, which is a typical characteristic of geographic SMR.
While the errors in estimating the communication bandwidth affect the state transfer time of the proposed method, they are not expected to be significant in a real environment.
This will be confirmed using an actual environment in the next section.

\section{Performance Evaluation}
\label{sec:evaluation}

Here, we build a geographic BFT-SMR system on Amazon EC2 using the open-source SMR library BFT-SMaRt 1.2\footnote{\url{https://github.com/bft-smart/library/releases/tag/v1.2}} and evaluate the state transfer time of the proposed method.

\subsection{Experimental Method}
We implement the proposed method and the CST protocol \cite{Bessani2013} in BFT-SMaRt as a state transfer method.
While the authors in \cite{Bessani2013} introduced various optimization techniques for the CST protocol, we implement only ``Optimizing CST,'' which reduces the state transfer time. 
In the CST protocol, the sizes of the checkpoint and logs sent by the transfer replicas depend on the timing of the state transfer, but we assume that each transfer replica sends the partial state of the same size because this is the fastest situation for the CST protocol.

In addition, as a baseline, we implement a method that removes the dynamic bandwidth estimation function from the proposed method.
This method allocates chunks using the average communication bandwidth measured in advance as follows.
The communication bandwidth between replicas is measured hourly for seven days (Worldwide Group is from April 26 to May 3, 2021, and European Group is from June 1 to June 8, 2021) using \texttt{iperf} 3.1.3.
Table \ref{tb:bandwidth} shows the measured communication bandwidth between each replica in two groups. 
In each table, the first column represents sending replicas and the first row represents receiving replicas.
Hereafter, we refer to this method as ``Premeasured BW.''

\begin{table}[t]
    \centering
    \caption{Premeasured inter-region communication bandwidth (Mbps).}
    \label{tb:bandwidth}
    \subfloat[Worldwide Group]{
        \small
        \begin{tabular}{lrrrr}
            \hline
            & Sydney & São Paulo & N. Virginia & Ireland \\ \hline
            Sydney &    & 33.7  & 57.0  & 42.9 \\
            São Paulo & 33.3   &   & 102.2  & 64.5 \\
            N. Virginia & 56.6   & 103.0  &   & 174.3 \\
            Ireland & 42.9   & 64.4  & 173.3  & \\
            \hline
        \end{tabular}
        \label{tb:bandwidth-a}
    }
    \quad
    \subfloat[European Group]{
        \small
        \begin{tabular}{lrrrr}
            \hline
            & Ireland & London & Paris & Frankfurt \\ \hline
            Ireland &    & 857.8  & 578.1  & 420.8 \\
            London & 866.8   &   & 1219.6  & 667.2 \\
            Paris & 594.4   & 1234.4  &   & 1115.5 \\
            Frankfurt & 420.0   & 662.4  & 1113.6  & \\
            \hline
        \end{tabular}
    
        \label{tb:bandwidth-b}
    } 

\end{table}

In the experiment, we use four replicas: one recovery replica and three transfer replicas.
The geographic locations of these replicas are Worldwide Group and European Group, described in Sect.~\ref{sec:introduction}.
Each replica uses a t3.xlarge instance.\footnote{\url{https://aws.amazon.com/ec2/instance-types/t3/}}
Table \ref{tb:aws-spec} shows the performance of the instance type.
In each replica, we use Docker 18.09.2 on Amazon Linux 2 to run BFT-SMaRt in the \texttt{openjdk:14-alpine container}. 
Unless noted otherwise, the size of the state to be transferred is 1000 MiB, the total number of chunks $N$ in the proposed method is 256, and the update interval $I$ of the chunk allocation in the proposed method is 1000 ms.
For the state transfer time, we use the average value of five measurements.
In the experiment, the size of the checkpoints is fixed, and each method divides only the checkpoint.
The log size is less than 1/1000 of the fixed state size in the preliminary experiment, and, thus, does not affect the result.
Therefore, we refer to the checkpoint size as the \emph{state size}.
We use SHA-512 as a hash function $\mathrm{Hash}(x)$ to verify the integrity of the chunks or the checkpoint and logs.

\begin{table}[t]
    \centering
    \caption{Specifications of t3.xlarge instance.}
    \label{tb:aws-spec} %
    \small
    \begin{tabular}{lr}
        \hline
        Item & Specification \\
        \hline
        CPU & Intel Xeon Platinum 8000 series \\
        \#vCPUs & 4 \\
        Memory & 16.0 GiB \\
        Network burst bandwidth & 5 Gbps\\
        \hline
    \end{tabular}
\end{table}

\subsection{Comparison of State Transfer Time}
\label{sec:compare-transfer-time}

The state transfer time of the proposed method is compared with that of the method using the CST protocol and premeasured BW.
We build geographic SMR systems for each group (the Worldwide Group and the European Group) and measure the state transfer time 12 times every 2 hours using each replica in the groups as a recovery replica and obtain the average value.
The experiment was conducted from June 18 to June 19, 2021.

\begin{figure}[t]
    \centering
    \subfloat[Worldwide Group]{
        \includegraphics[width=70mm]{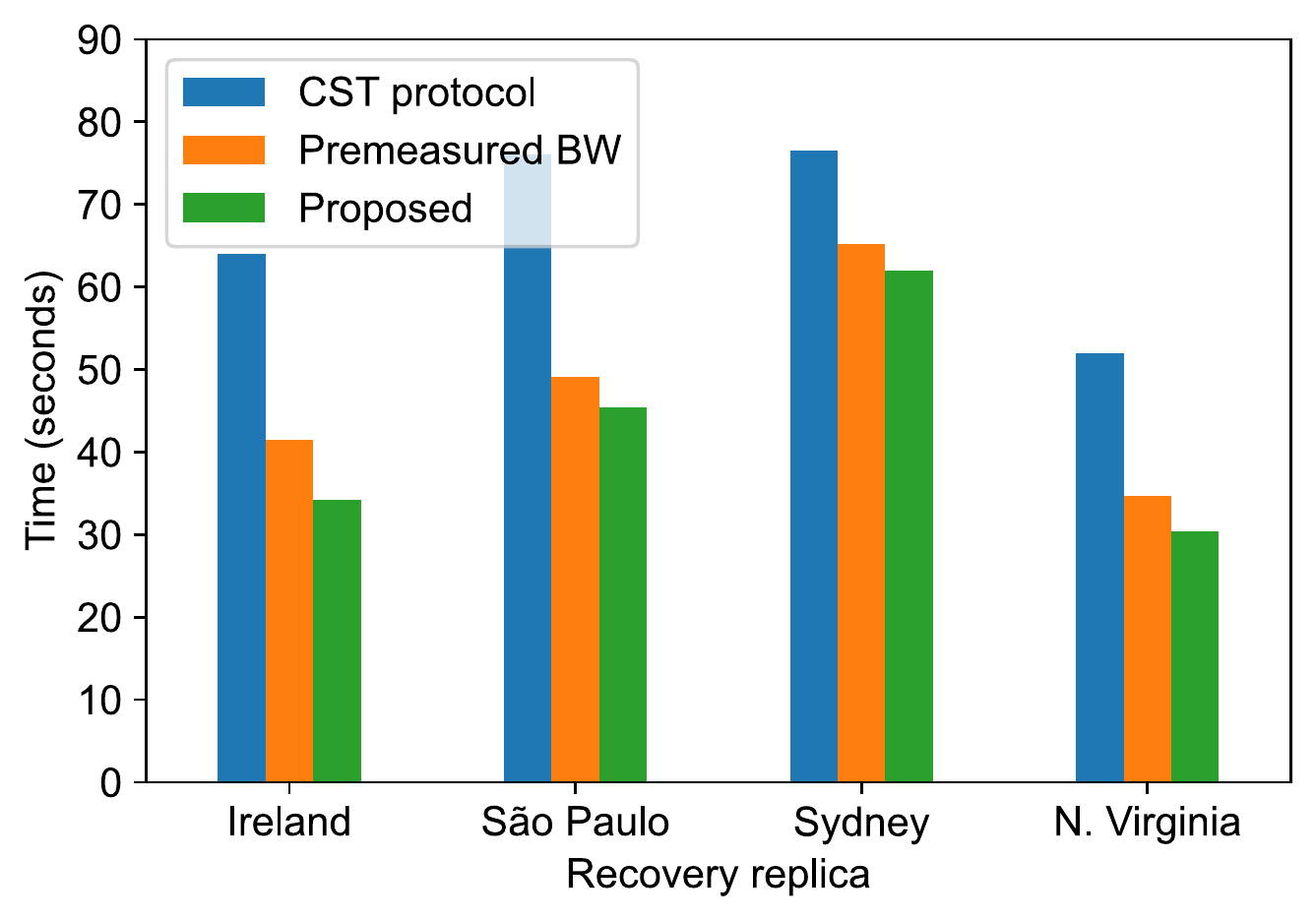}
        \label{fig:recovery-replica-and-transfer-time-group-a}
    }
    \subfloat[European Group]{
        \includegraphics[width=70mm]{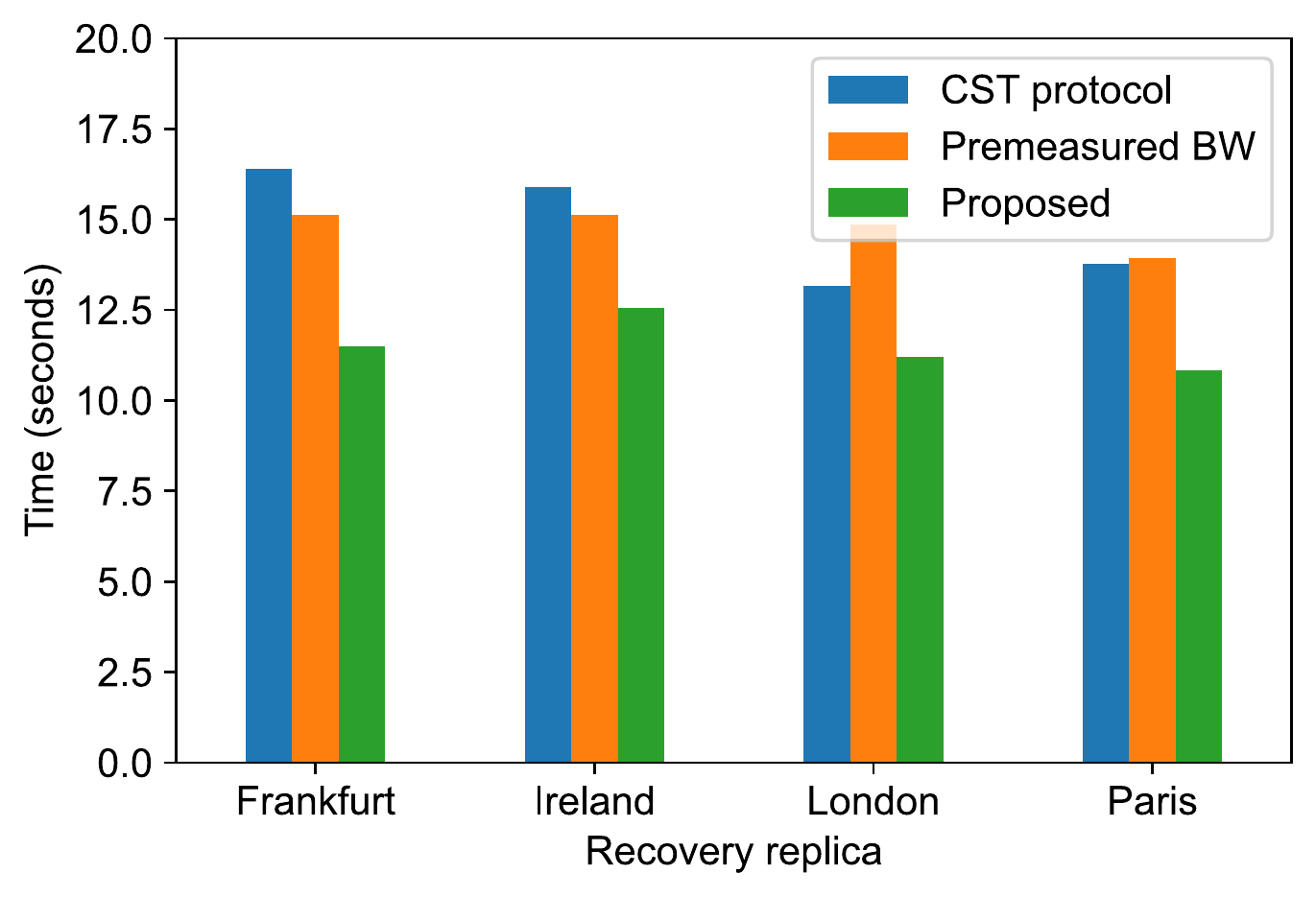}
        \label{fig:recovery-replica-and-transfer-time-group-b}
    }
    \caption{Location of the recovery replica and state transfer time when the state size is 1000 MiB.}
    \label{fig:recovery-replica-and-transfer-time}
\end{figure}

First, we compare the average state transfer time for each recovery replica when the state size is 1000 MiB.
Figure \ref{fig:recovery-replica-and-transfer-time} shows the results.
In the Worldwide Group, the state transfer time of the proposed method was the shortest for all recovery replicas.
When the recovery replica was in Ireland, the difference between the proposed method and the CST protocol was the largest and reduced by 47\%.
In contrast, the smallest reduction to the CST protocol (i.e., 19\%) occured, when the recovery replica was in Sydney.
This difference is due to the difference in the communication bandwidth of the transfer replicas.
The average reduction rate of the proposed method for the CST protocol was 37\%.
In contrast, the average reduction rate of the proposed method to the premeasured BW method was 10\%.
This is because the time variation of the communication bandwidth between the replicas is small.

Unlike Worldwide Group, the state transfer time of the premeasured BW method was longer in the European Group.
This is because, as shown in Fig.~\ref{fig:bandwidth-change}\subref{fig:bandwidth-change-group-b-london}, the time variation of the communication bandwidth is large in the European Group.
The results obtained for the European Group suggest the importance of dynamically responding to the time variation of the communication bandwidth.

\begin{figure}[t]
    \centering
    \subfloat[North Virginia (Worldwide Group)]{
        \includegraphics[width=70mm]{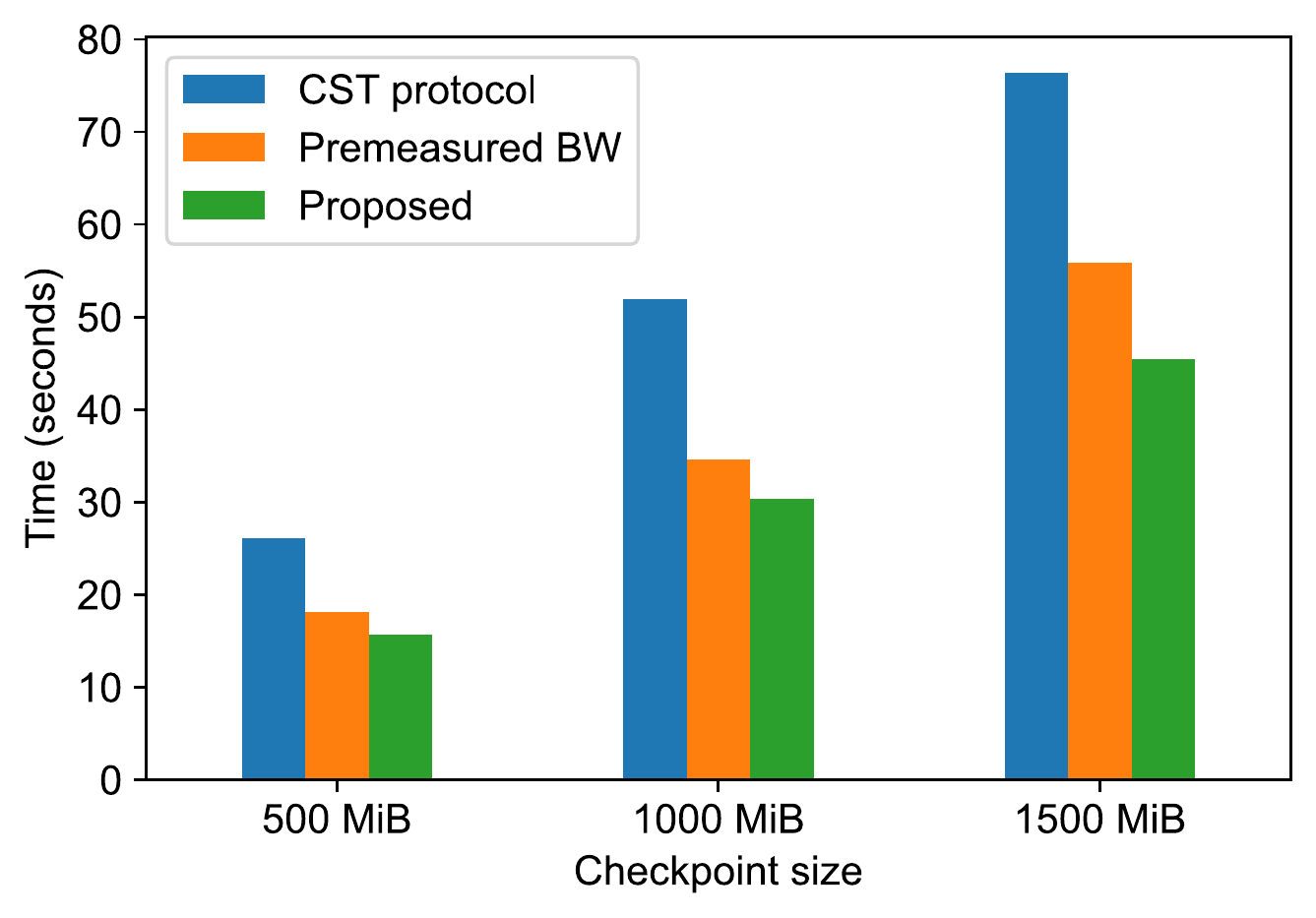}
        \label{fig:state-size-and-transfer-time-group-a}
    }
    \subfloat[Ireland (European Group)]{
        \includegraphics[width=70mm]{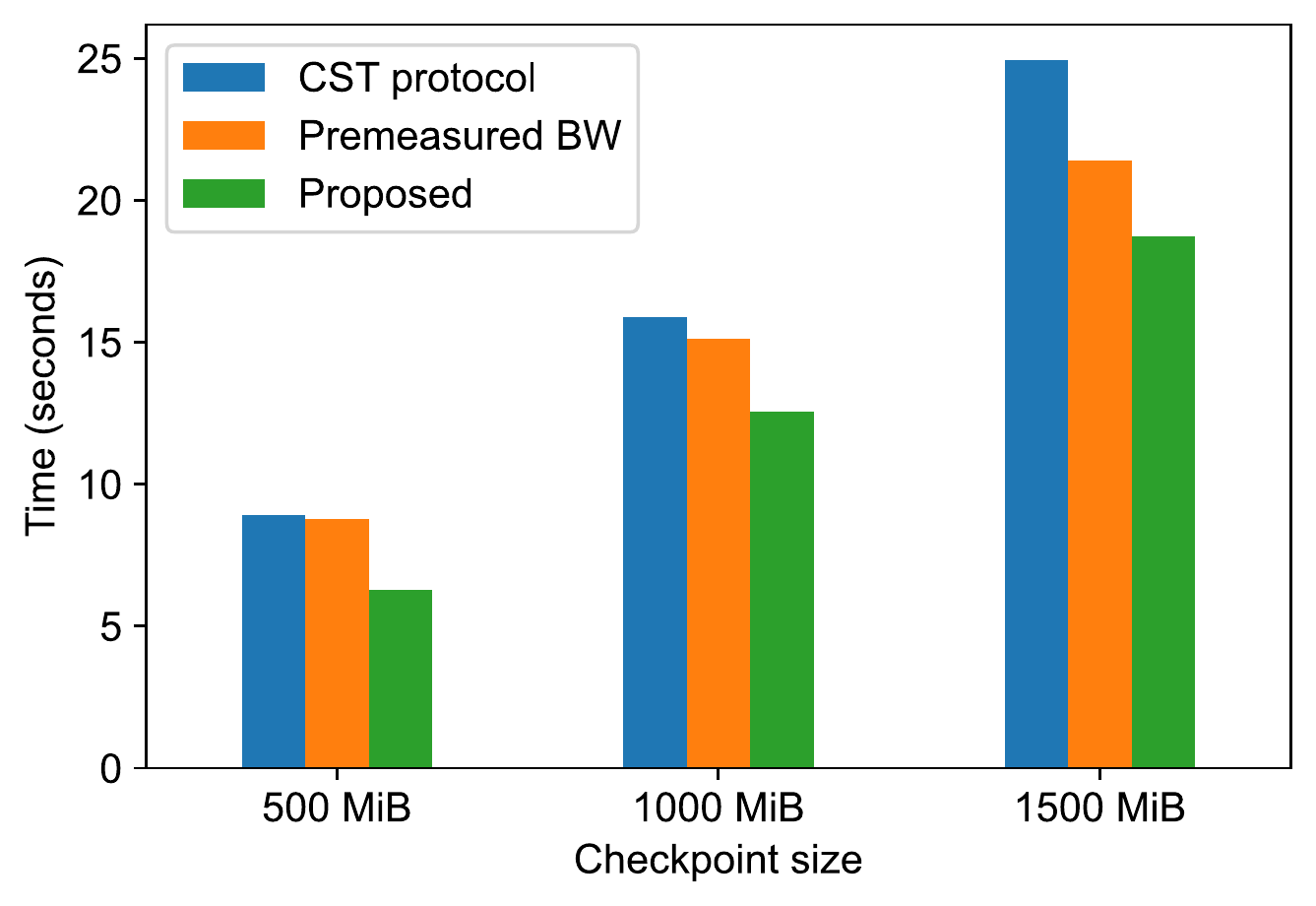}
        \label{fig:state-size-and-transfer-time-group-b}
    }
    \caption{State size and state transfer time.}
    \label{fig:state-size-and-transfer-time}
\end{figure}

Next, we compare the effects of different state sizes on the state transfer time.
Figure \ref{fig:state-size-and-transfer-time} shows the average state transfer times for different state sizes of 500, 1000, and 1500 MiB in North Virginia in the Worldwide Group and Ireland in the European Group. 
In the Worldwide Group, the effects of the state sizes were small: 40\% for the 500 MiB state, 42\% for the 1000 MiB state, and 41\% for the 1500 MiB state, compared to the CST protocol.
In contrast, in the European Group, the effects of state sizes were larger than those in the Worldwide Group: 30\% for the 500 MiB state, 21\% for the 1000 MiB state, and 25\% for the 1500 MiB state.
This is because the European Group was more strongly affected by the increase in transfer time due to the larger state size since the time variation in the communication bandwidth was larger than that in the Worldwide Group.

\subsection{State Transfer Time of Each Transfer Replica}
\label{sec:transfer-time-for-each-transfer-replicas}

\begin{figure}[t]
    \centering
    \subfloat[North Virginia (Worldwide Group)]{
        \includegraphics[width=70mm]{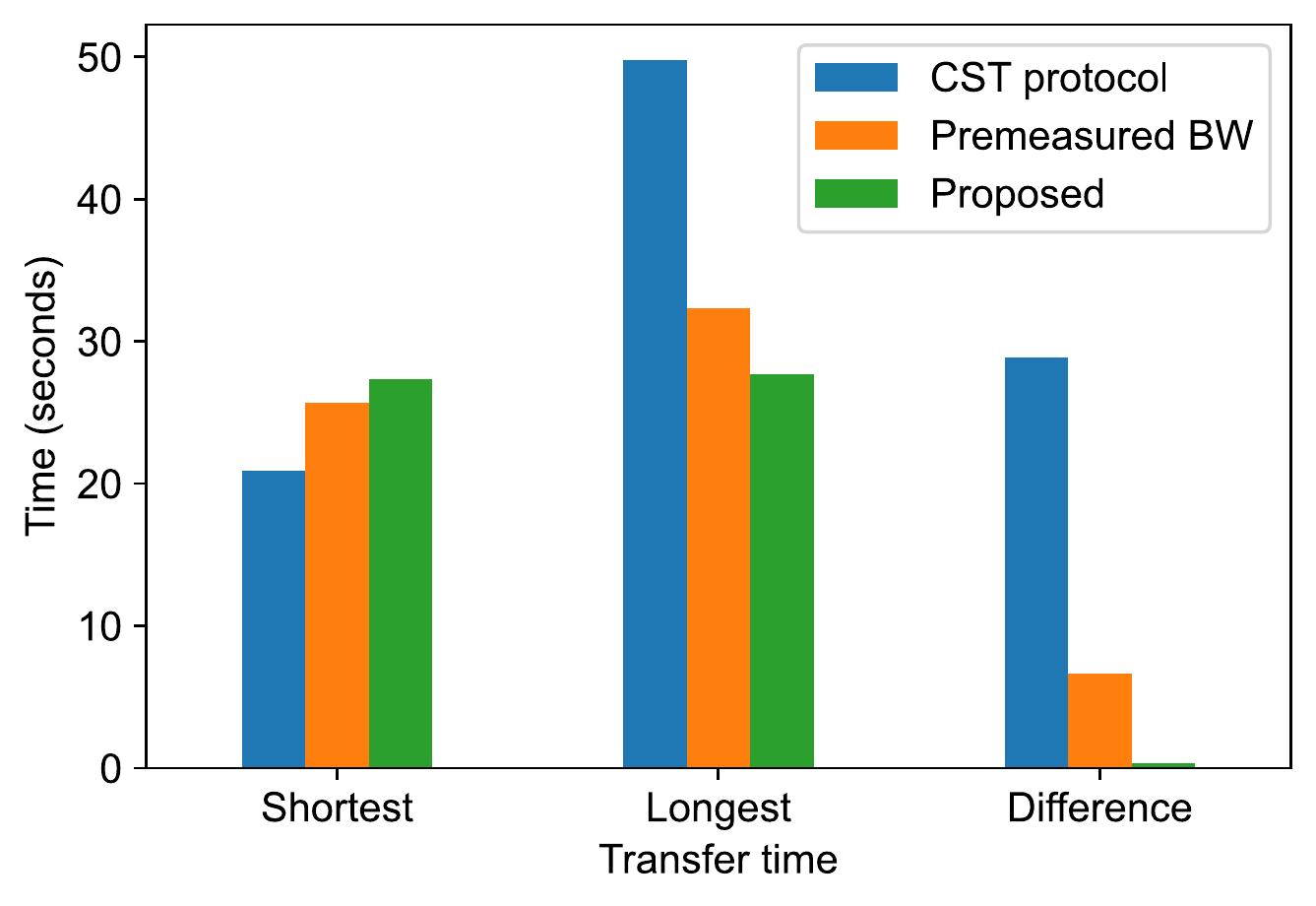}
        \label{fig:time-per-transfer-steps-a-virginia}
    }
    \subfloat[Ireland (European Group)]{
        \includegraphics[width=70mm]{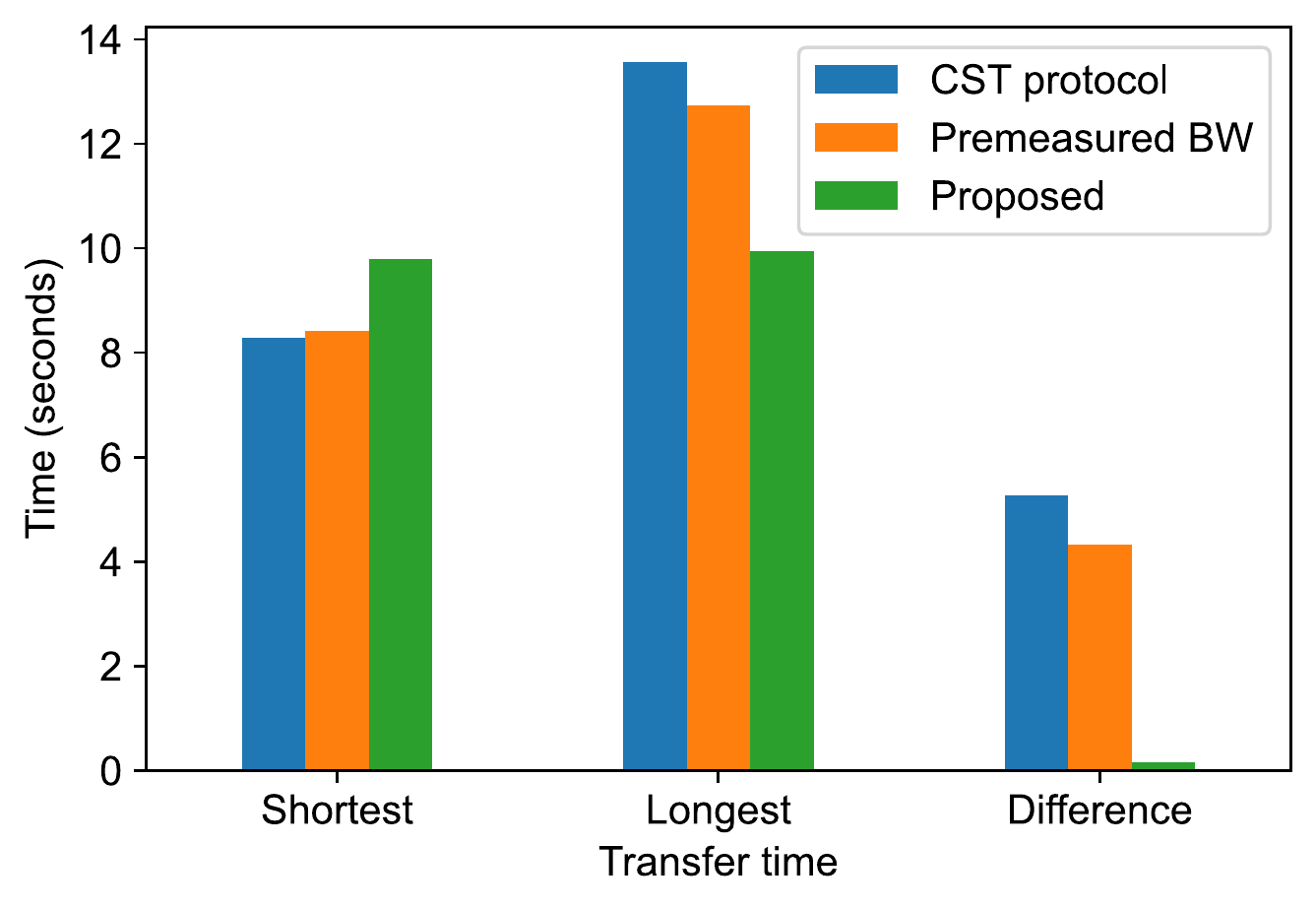}
        \label{fig:time-per-transfer-steps-b-ireland}
    }
    \caption{State transfer time for each replica. The horizontal axis shows the shortest and longest transfer times of the transfer replicas and their difference.}
    \label{fig:time-per-transfer-steps}
\end{figure}

The proposed method reduces the total state transfer time by balancing the state transfer time of the transfer replicas.
We demonstrate the effect of balancing the state transfer time by comparing the state or chunk transfer times of the transfer replicas.
This experiment uses North Virginia in the Worldwide Group and Ireland in the European Group as the recovery replicas.
The experiment was conducted from June 18 to June 19, 2021.
Figure \ref{fig:time-per-transfer-steps} shows the state transfer time for each transfer replica.

As shown in Fig.~\ref{fig:time-per-transfer-steps}\subref{fig:time-per-transfer-steps-a-virginia} and Fig.~\ref{fig:time-per-transfer-steps}\subref{fig:time-per-transfer-steps-b-ireland}, the CST protocol and the premeasured BW method have large difference in the state transfer times of the transfer replicas.
The recovery replicas in North Virginia and Ireland have differences of 2.38 and 1.63 times with the CST protocol and 1.26 and 1.54 times with the premeasured BW, respectively.
The CST protocol has no mechanism to adjust the difference in the communication bandwidth between the replicas.
Therefore, the state transfer time tends to differ considerably for each replica.
This tendency is particularly remarkable in the Worldwide Group.
The premeasured BW method cannot follow the changes in the communication bandwidth while transferring chunks.
Therefore, the state transfer time tends to vary for each transfer replica in the European Group, where the communication bandwidth changes frequently.
In contrast, the difference in the proposed method is significantly small, 1.01 times in both cases.
The proposed method has features of dynamically adapting to both problems, and it works well.

\subsection{Time Variation of Chunk Allocation of the Proposed Method}
\label{sec:temporal-change-in-chunk-allocate}

\begin{figure}[t]
    \centering
    \includegraphics[width=70mm]{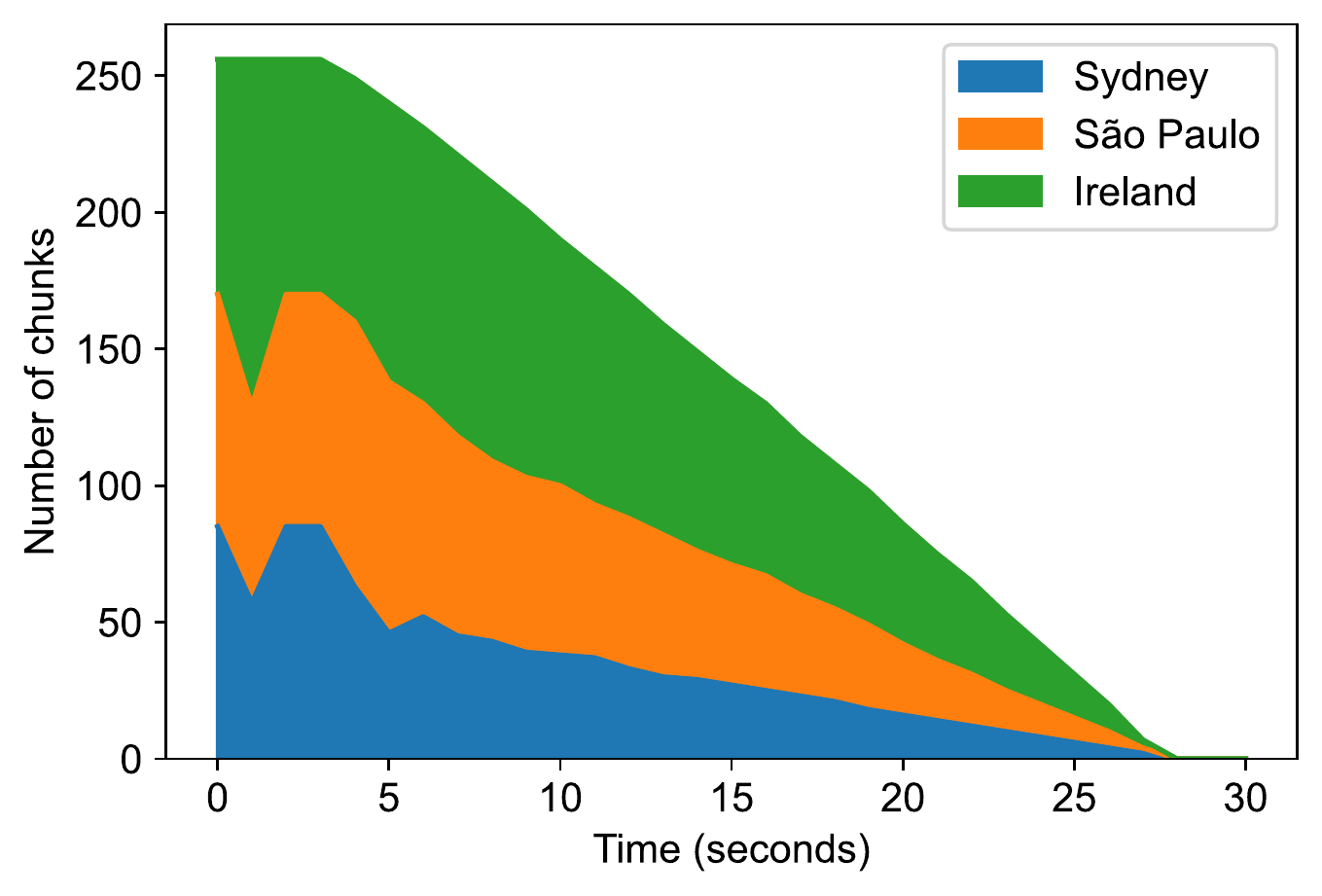}
    \caption{Time variation of chunk allocation (North Virginia, Worldwide Group).}
    \label{fig:temporal-change-group-a-virginia}
\end{figure}

Here, we evaluate the time variation of chunk allocation of the proposed method to verify that the method appropriately allocates chunks to each transfer replica based on the communication bandwidths.
In this evaluation, the state transfer is repeated five times, and we select the trial with the median transfer time.
The experiment was conducted on August 12, 2021.
Figure \ref{fig:temporal-change-group-a-virginia} shows the time variation of chunk allocation when the recovery replica is in North Virginia in the Worldwide Group.

As shown in Fig.~\ref{fig:temporal-change-group-a-virginia}, the number of remaining chunks changed for the first 4 seconds after the start of the transfer.
This indicates that the recovery replica has not received chunks for more than 3 seconds after the start.
The change in the chunk allocation for these 4 seconds is caused by not having enough data to be transferred to estimate the communication bandwidth because the chunk reception has not started.
After the 4 seconds, chunks are allocated at an almost constant rate, and the requested chunks decrease at a constant slope.
This indicates that the transfer rate of the proposed method is close to constant, and the correct communication bandwidth is reflected in the chunk allocation.
Figure \ref{fig:temporal-change-group-a-virginia} shows that the time to start the state transfer is more than 3 seconds.
Although each transfer replica calculates hashes before sending chunks, the calculation time is approximately 2.3 seconds in this experiment.
Since the hash calculation time is shorter than the time to start the transfer, the delay in sending chunks cannot be explained by the hash calculation time alone.
Therefore, we suppose that the delay in the start of the transfer was caused by two factors: the hash calculation time and the latency of the transfer replica.

\subsection{Effects of Different Hash Calculation Methods on State Transfer Time}
\label{sec:relation-between-hash-calc-way-and-transfer-time}

\begin{figure}[t]
    \centering
    \includegraphics[width=70mm]{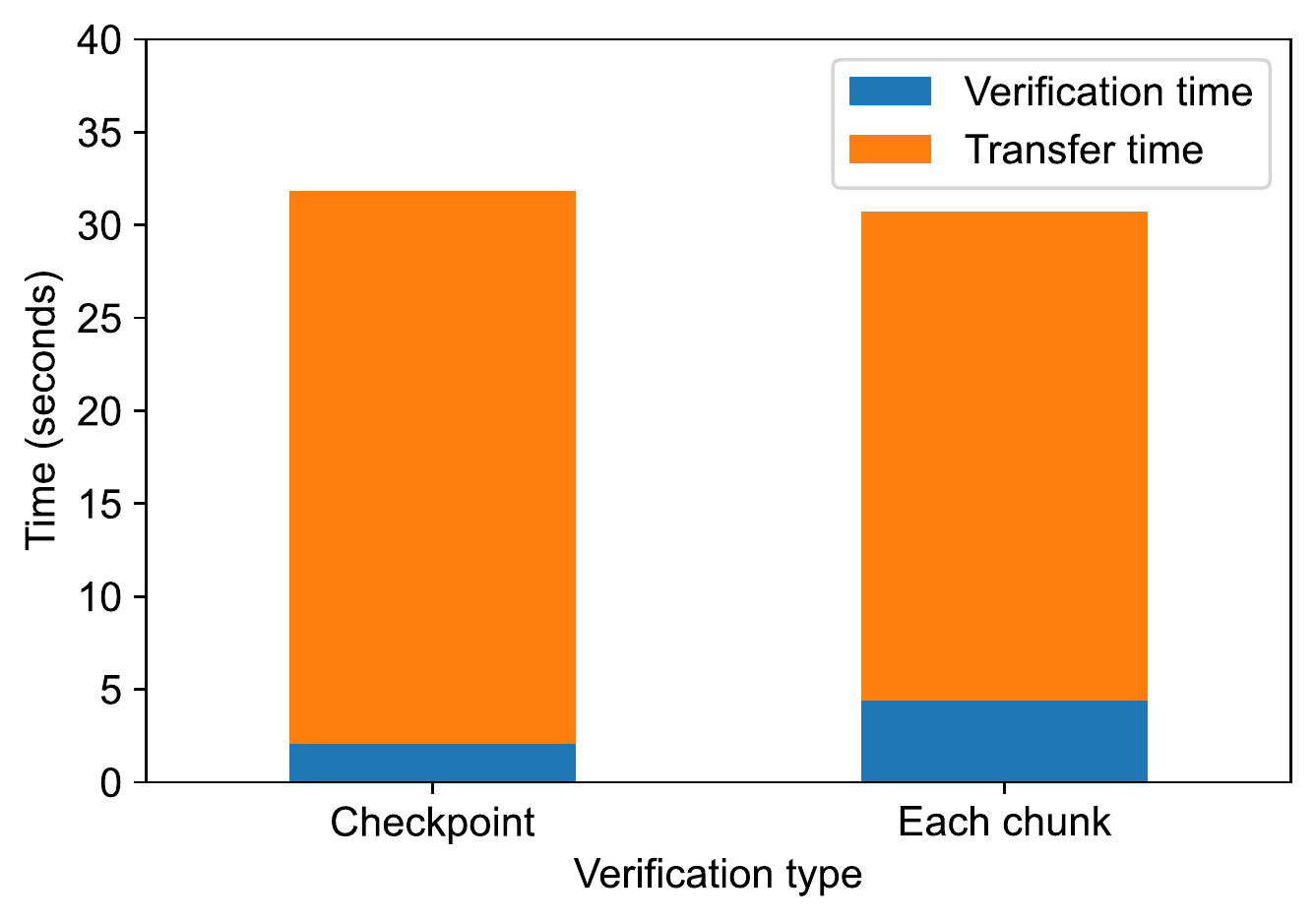}
    \caption{State transfer time for each hash verification method (North Virginia, Worldwide Group). The state transfer time includes the verification time.}
    \label{fig:relation-between-hash-calc-way-and-transfer-time-group-a-virginia}
\end{figure}

In the proposed method, a recovery replica calculates the hash of each chunk and verifies the hash in parallel with the chunk reception to reduce the transfer time.
To show the effect of this technique, we measure the difference in transfer time between the proposed method and the method that calculates the hash of the whole state.
In the latter method, all transfer replicas send the whole hash at the start of the transfer, and the recovery replica verifies the hash after the chunks are combined.
The experiment was conducted from June 9 to 10, 2021.
Figure \ref{fig:relation-between-hash-calc-way-and-transfer-time-group-a-virginia} shows the results measured by a recovery replica in North Virginia in the Worldwide Group.

Figure \ref{fig:relation-between-hash-calc-way-and-transfer-time-group-a-virginia} shows that the proposed method increases the hash verification time but decreases the total transfer time.
The increase in hash verification time is caused by the overhead of frequent hash function calls and initialization caused by the division of chunks.
The decrease in transfer time was about half the hash verification time of the method calculating the entire hash, indicating that about half of the hash verification time could be executed in parallel during the state transfer.

\subsection{Effect of Update Interval of the Chunk Assignment}
\label{sec:chunk-num-frequency-and-transfer-time}

\begin{figure}[t]
    \centering
    \includegraphics[width=70mm]{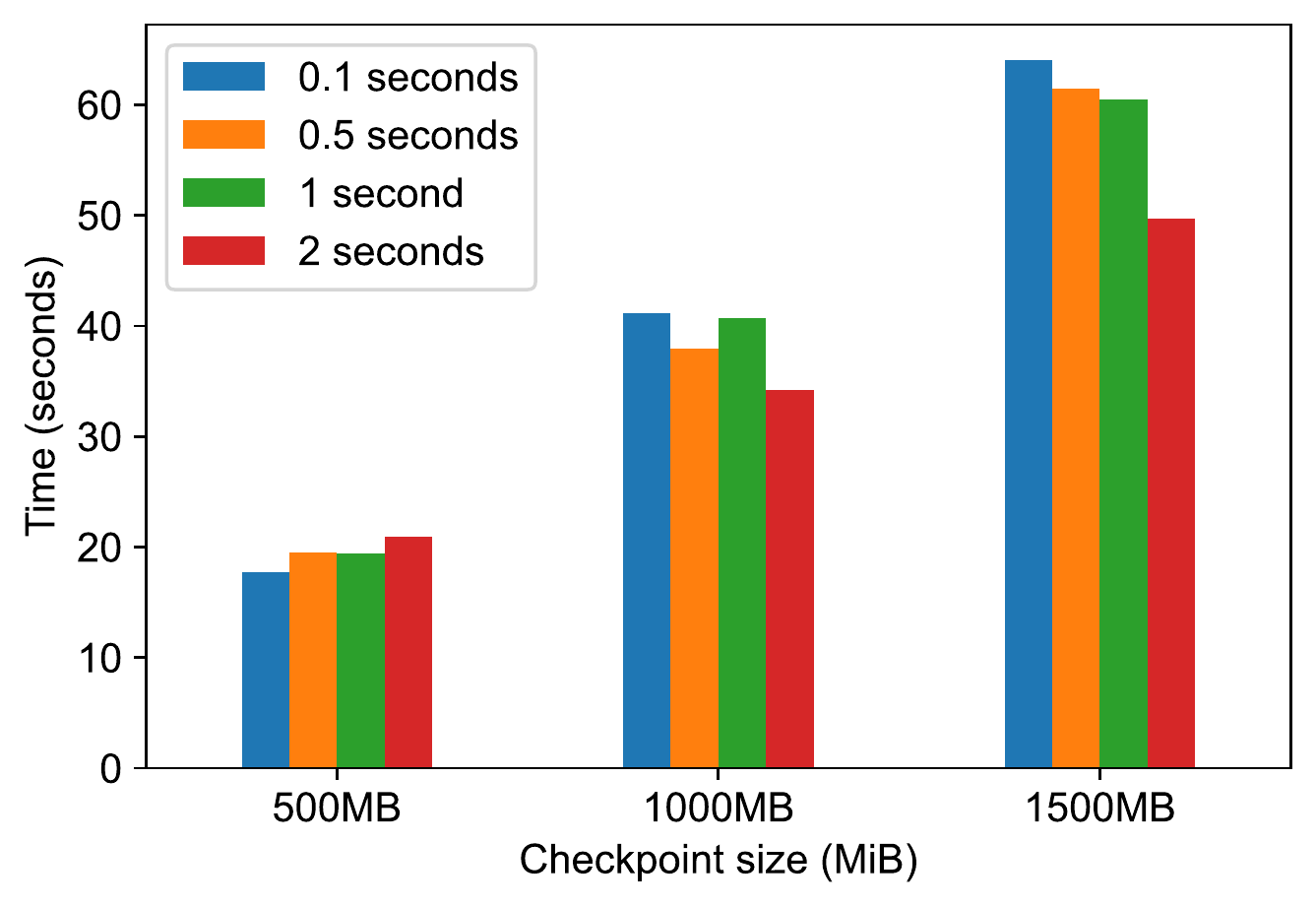}
    \caption{Chunk update interval and state transfer time (Worldwide Group, the recovery replica is in North Virginia).}
    \label{fig:chunk-num-frequency-and-transfer-time-group-a-virginia}
\end{figure}

In the proposed method, the recovery replica responds to changes in the communication bandwidth over time by repeatedly updating the number of chunks that the transfer replicas send.
To verify the effect of the chunk update interval $I$ on the state transfer time, we measure the state transfer time by varying $I$ as 0.1,  0.2, 0.5, 1, and 2 seconds.
The experiment was conducted on August 12, 2021.
The results are shown in Fig.~\ref{fig:chunk-num-frequency-and-transfer-time-group-a-virginia}, where the replicas are placed in the Worldwide Group, and the recovery replica is in North Virginia.
 
As shown in Fig.~\ref{fig:chunk-num-frequency-and-transfer-time-group-a-virginia}, when the state size is 500 MiB, the shorter the update interval is, the shorter the transfer time is. 
However, when the state size is 1000 MiB or larger, the longer the update interval is, the shorter the transfer time is.
This trend can be explained by the transition of chunk allocation shown in Sect.~\ref{sec:temporal-change-in-chunk-allocate}.
First, when the state size is 500 MiB, the state transfer time is always short.
Since the communication bandwidth is unstable for a relatively long time at the beginning of the state transfer,
the proposed method with the shorter update interval could estimate the bandwidth accurately.
In contrast, when the state size is 1000 MiB or more, the state transfer time is longer than that at the state size of 500 MiB.
Since the communication bandwidth remains stable for a long time, the method with a longer update interval can reduce the impact of updating the chunk allocation.

\subsection{Relationship between the Number of Chunks and State Transfer Time}
\label{sec:total-chunk-num-and-transfer-time}

\begin{figure}[t]
    \centering
    \includegraphics[width=70mm]{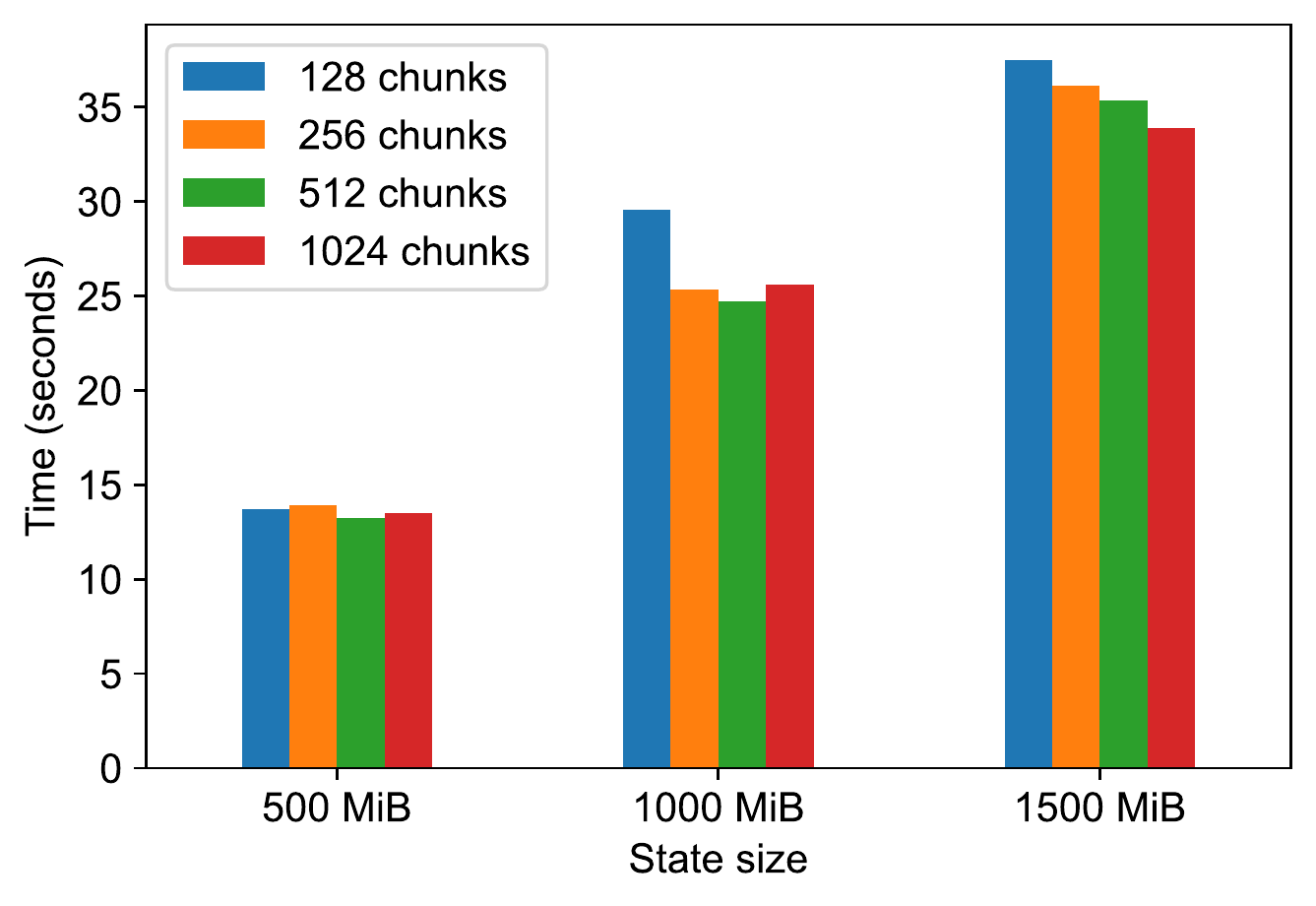}
    \caption{Number of chunks $N$ and state transfer times (Worldwide Group, the recovery replica is in North Virginia).}
    \label{fig:total-chunk-num-and-transfer-time-group-a-virginia}
\end{figure}

The proposed method divides the service state into $N$ chunks and transfers them to the recovery replica.
Here, we investigate how the number of chunks affects the state transfer time and vary $N$ as 128, 256, 512, and 1024.
The experiment was conducted on March 8, 2022.
The results are shown in Fig.~\ref{fig:total-chunk-num-and-transfer-time-group-a-virginia}, where the replicas are placed in the Worldwide Group, and the recovery replica is in North Virginia.

As shwon in Fig.~\ref{fig:total-chunk-num-and-transfer-time-group-a-virginia}, the state transfer time decreases as the number of chunks $N$ is increased, although the details differ depending on the state size.
This tendency is remarkable when the state size is 1500 MiB, and the state transfer time is 9.6 \% shorter when $N = 1024$ compared to the case where $N = 128$.
However, there is a threshold above which increasing $N$ does not shorten the state transfer time.
For instance, we can see the threshold at $N = 256$ when the state size is 1000 MiB.

\subsection{Error Rate in Estimated Communication Bandwidth during State Transfer}
\label{sec:bandwidth-estimation-error}

\begin{figure}[t]
    \centering
    \includegraphics[width=70mm]{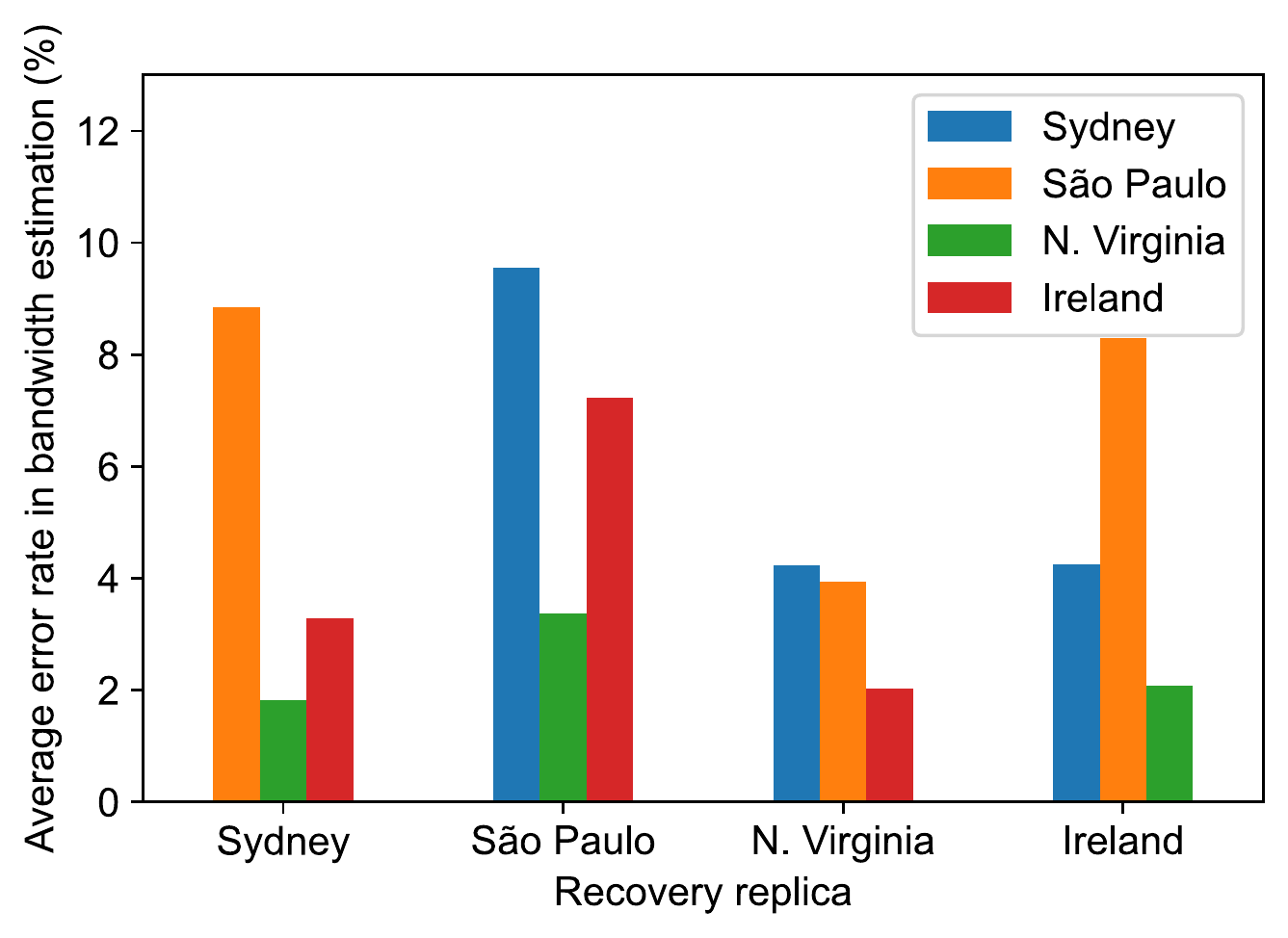}
    \caption{Error rate between the estimated bandwidth and actual bandwidth during state transfer (Worldwide Group).}
    \label{fig:bandwidth-estimation-error}
\end{figure}

In the proposed method, a recovery replica estimates the communication bandwidth between itself and each transfer replica from the amount of data received during a state transfer.
Here, to verify the accuracy of this estimation, we measure the error rate between the estimated communication bandwidth and the actual communication bandwidth every $I$ second during the state transfer.
We repeat this measurement 12 times every 2 hours to obtain the final average value.
Here, we denote the communication bandwidth estimated from the amount of data received in the interval $ I_i $ by $ e_i $.
In this case, the error between the actual bandwidth and the estimated bandwidth in interval $I_i$ is expressed as $|e_{i+1} - e_i|$.
We measure the estimation error in the period from 5 seconds after the start of state transfer to 5 seconds before the end of the state transfer, because the communication bandwidth is stable in this period.
This experiment was conducted from June 18th to June 19th, 2021.
The result obtained at a state size of 1000 MiB in the Worldwide Group is shown in Fig.~\ref{fig:bandwidth-estimation-error}.

Fig.~\ref{fig:bandwidth-estimation-error} shows that the error between the actual communication bandwidth and the estimated bandwidth is within 10\%.
The analysis in Sect.~\ref{sec:analysis} shows that the proposed method can transfer the state faster than the existing methods when the error is within 10\%, and we can confirm this fact from the experimental results.
In addition, unlike the analysis shown in Sect.~\ref{sec:analysis}, in the actual environment, the case of overestimating the actual bandwidth (which increases the transfer time) and the case of underestimating the actual bandwidth (which decreases the transfer time) cancel each other out.
Therefore, the effect on state transfer is even smaller.

\section{Applying the Proposed Method to Dynamic Replica replacement}
\label{sec:replica-relocation}

In a distributed system consisting of multiple geographically-separated servers, the latency of a system changes depending on the geographical arrangement of the servers.
For this reason, the authors in \cite{Karve2006,Agarwal2010} proposed a method to optimize the system latency by moving servers and the data stored on them according to changes in latency between the servers.
It is also known that the latency of a geographic SMR system also changes depending on the geographical arrangement of replicas \cite{Numakura2019b,Sousa2015}.
Since the latency of the geographic SMR system is determined from the latency between the replicas, moving replicas\footnote{While we use the term ``move a replica'' here, a replica is moved in two steps. First, we place a new replica at a destination. After the state transfer to the replica is complete, the original replica is deleted from replication.} can reduce the system latency.
Here, we call this approach \emph{dynamic replica replacement}.
Dynamic replica replacement in geographic SMR has several advantages: the mechanism is simpler than the methods using additional replicas \cite{Sousa2015,Berger2019}, and the installation cost of additional replicas can be reduced.
However, since the state transfer, which is required for moving a replica, is considered a time-consuming process \cite{Sousa2015}, dynamic replica replacement has not been attempted so far.
Since the proposed method can transfer a service state fast as we observed in Sections \ref{sec:analysis} and \ref{sec:evaluation}, we expect that the performance degradation of dynamic replica replacement also can be alleviated.
Therefore, we evaluate the practicality of dynamic replica replacement using the proposed method.

\subsection{Evaluation Settings}

We assume the following two scenarios that require dynamic replica replacement and reproduce them.
\begin{itemize}
	\item Scenario 1: Replace a replica with poor latency between replicas due to a failure of replica or network with a close replica
	\item Scenario 2: Improve system latency by moving a replica with high latency between replicas to a different location
\end{itemize}
Here, we call the replica to be removed to improve the system latency \emph{removal replica}, and that to be added to a system instead of the removal replica \emph{additional replica}.

In the experiment, we construct a geographic BFT-SMR system using three transfer replicas, one removal replica, and one additional replica, for a total of five replicas.
The additional replica does not participate in the agreement during the state transfer and receives only the result of the agreement to prevent the latency from deteriorating due to the increase in the number of replicas.
The configuration of each replica and each client is the same as that in Sect.~\ref{sec:evaluation}.
However, due to the restriction of the software used for this experiment\footnote{The \texttt{tc} command did not work in the docker container used in Sect.~\ref{sec:evaluation}.}, we use \texttt{openjdk:11-jre-slim-bullseye} as the docker container.
We have confirmed in advance that this change will not affect the system latency.
We use the latency from request transmission to response reception measured by a client as an evaluation measure.
In geographic SMR, the smaller the latency between the leader replica and a client is, the smaller the system latency is \cite{Sousa2015}.
Therefore, we place the client at the same location as the leader replica.
The client sends requests to each replica synchronously, and as soon as it receives the response, it sends the next request.
In this experiment, we compare two state transfer methods in dynamic replica replacement: the proposed method and PBFT state transfer.
Compared to the CST protocol, which was used in Sect.~\ref{sec:evaluation}, PBFT state transfer undergoes lesser performance degradation when the network bandwidth is limited due to a failure of replica or network assumed in the experiment.
Therefore, we use PBFT state transfer as a baseline here.
The state size to be transferred is 1000 MiB, the total number of chunks of the proposed method is $N = 256$, and the update frequency of the chunk assignment is $I = 1000$ ms.

\subsection{Replacing the Replica with Poor Latency between Replicas}
\label{sec:dynamic-relocation-when-replica-lantecy-becomes-bad}

\begin{figure}[t]
    \centering
    \includegraphics[width=75mm]{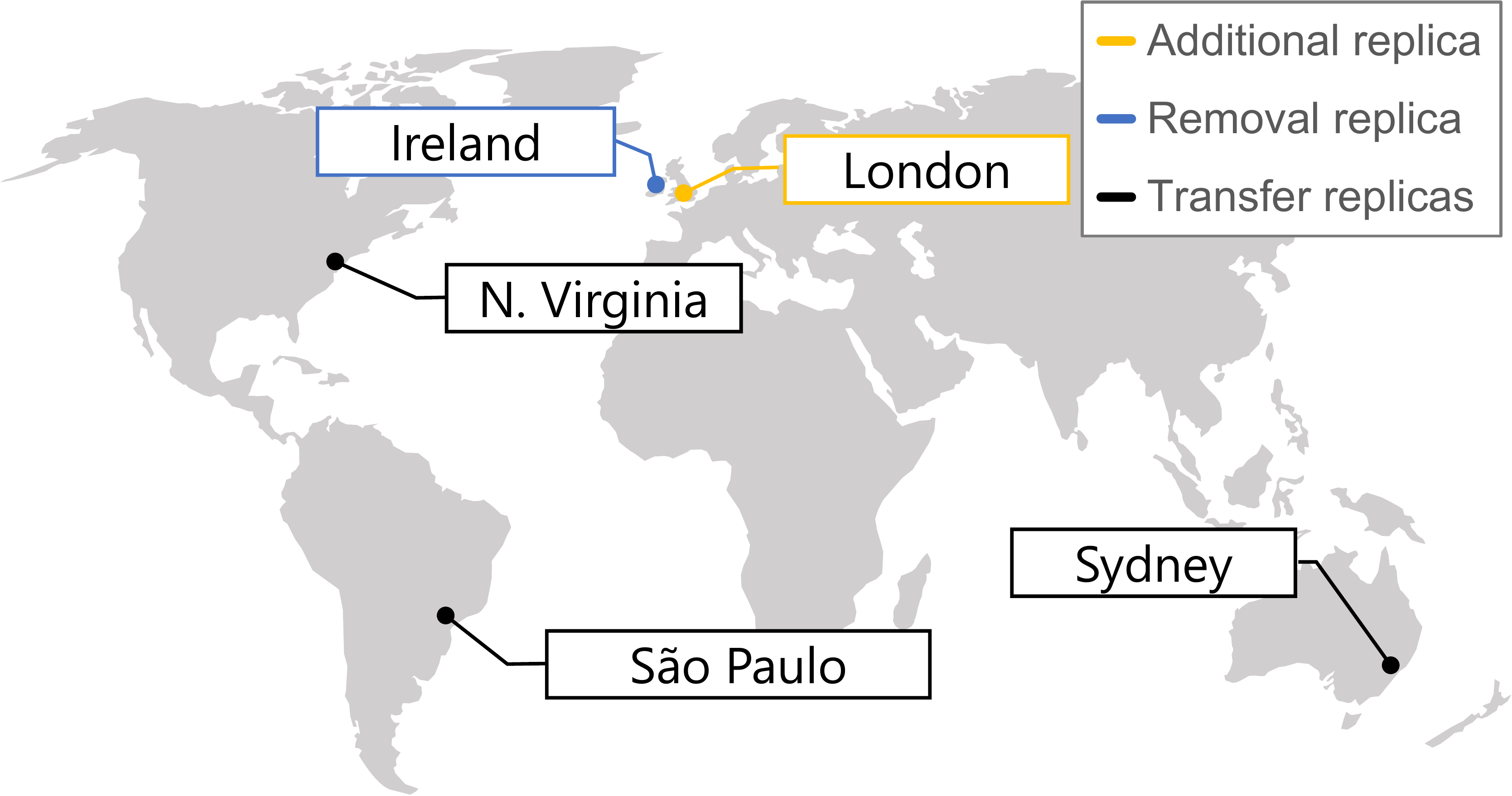}
    \caption{Replica deployment in the experiment of replica replacement.}
    \label{fig:relocation-map-london}
\end{figure}

Here, we assume Scenario 1, in which a replica with deteriorated latency between replicas is replaced with a replica close to the slow replica, and measure the change in latency of the geographic BFT-SMR system.
We worsen the latency between the removal replica and other replicas by artificially adding a delay to all messages sent by a removal replica with \texttt{tc} command\footnote{\url{https://man7.org/linux/man-pages/man8/tc.8.html}}.
The delay to be added is 100 ms.
Thirty seconds after the latency deteriorates, the additional replica is added to the system and begins state transfer.
After the state transfer of the additional replica is completed, the removal replica is removed from the system.
When a replica or a network fails, it is unrealistic for the removal replica to participate in the state transfer.
Therefore, in this experiment, the removal replica does not participate in the state transfer.
As shown in Fig.~\ref{fig:relocation-map-london}, replicas are placed in the Worldwide Group, defined in Sect.~\ref{sec:introduction}, and London.
The removal replica and the additional replica are in London and Ireland, respectively.
Both the leader replica and a client are in Sydney.
The experiment was conducted on February 3, 2022.

\begin{figure}[t]
    \centering
    \includegraphics[width=75mm]{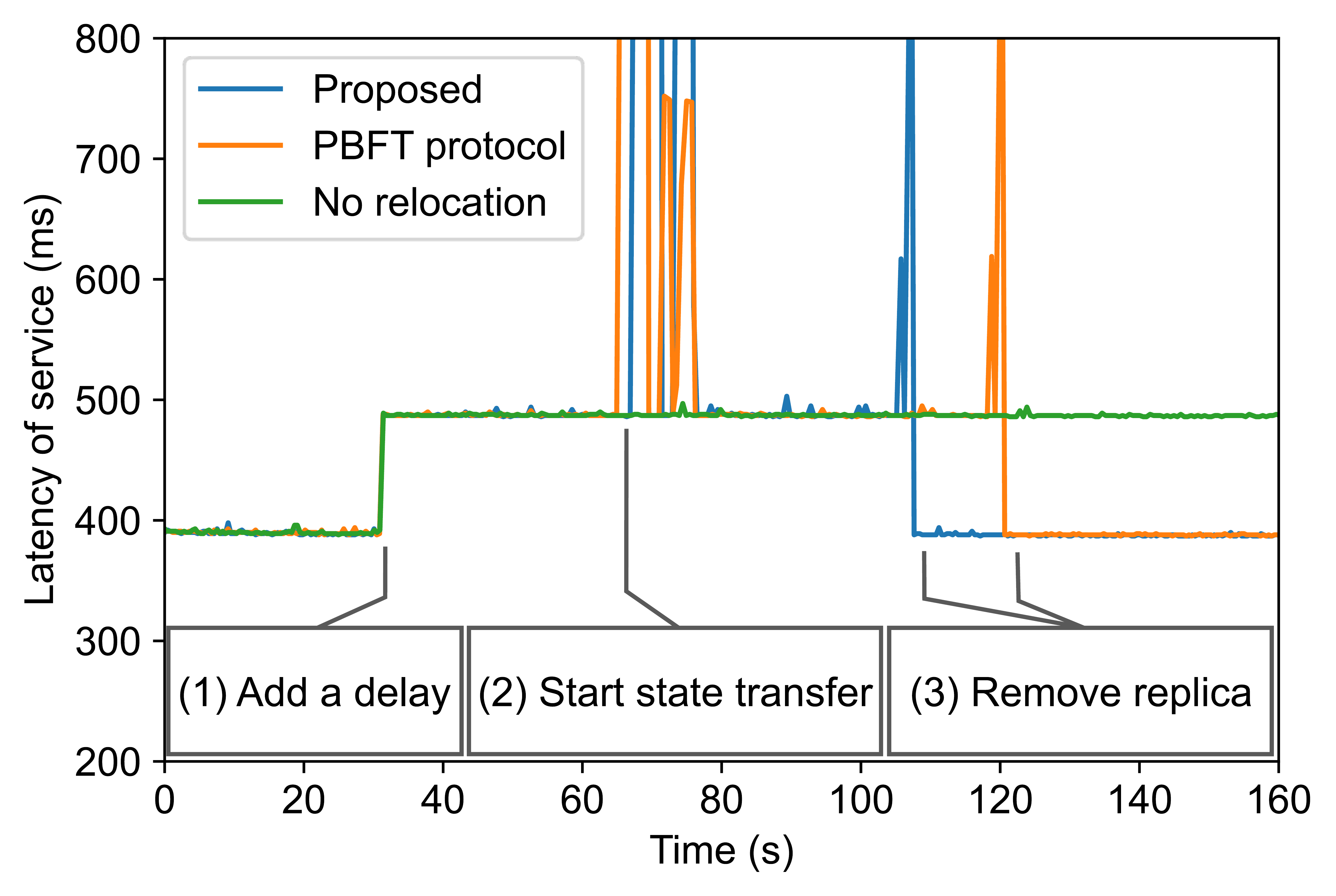} 
    \caption{Latency change in the experiment of replica replacement.}
    \label{fig:replica-relocation-london-to-ireland}
\end{figure}

The results are shown in Fig.~\ref {fig:replica-relocation-london-to-ireland}.
At (1) of Fig.~\ref{fig:replica-relocation-london-to-ireland}, a delay was added to the removal replica to simulate a failure of replica or network.
At this time, the system latency also increased by 100 ms, similar to the added delay.
At (2), the recovery replica started state transfer to the additional replica.
At this time, in both the proposed method and PBFT, the request processing was stopped for approximately one second.
This is because it is necessary to stop processing consensus temporarily to change the replication configuration (i.e., adding the additional replica).
At (3), when the status transfer to the additional replica was completed, the removal replica was deleted from the system.
Since London and Ireland are close to each other and the tendency of latency between replicas is similar, the system latency recovered to the same level as before (1).

The above results show that the deteriorated system latency can be recovered by replacing the replica whose latency has deteriorated with another replica by dynamic replica replacement.
While PBFT took 55 seconds for state transfer, the proposed method completed the state transfer in 40 seconds.
Thus, the proposed method is effective for replica replacement with poor latency between replicas.

\subsection{Migrating a Replica with Large Latency between Replicas}

\begin{figure}[t]
    \centering
    \includegraphics[width=75mm]{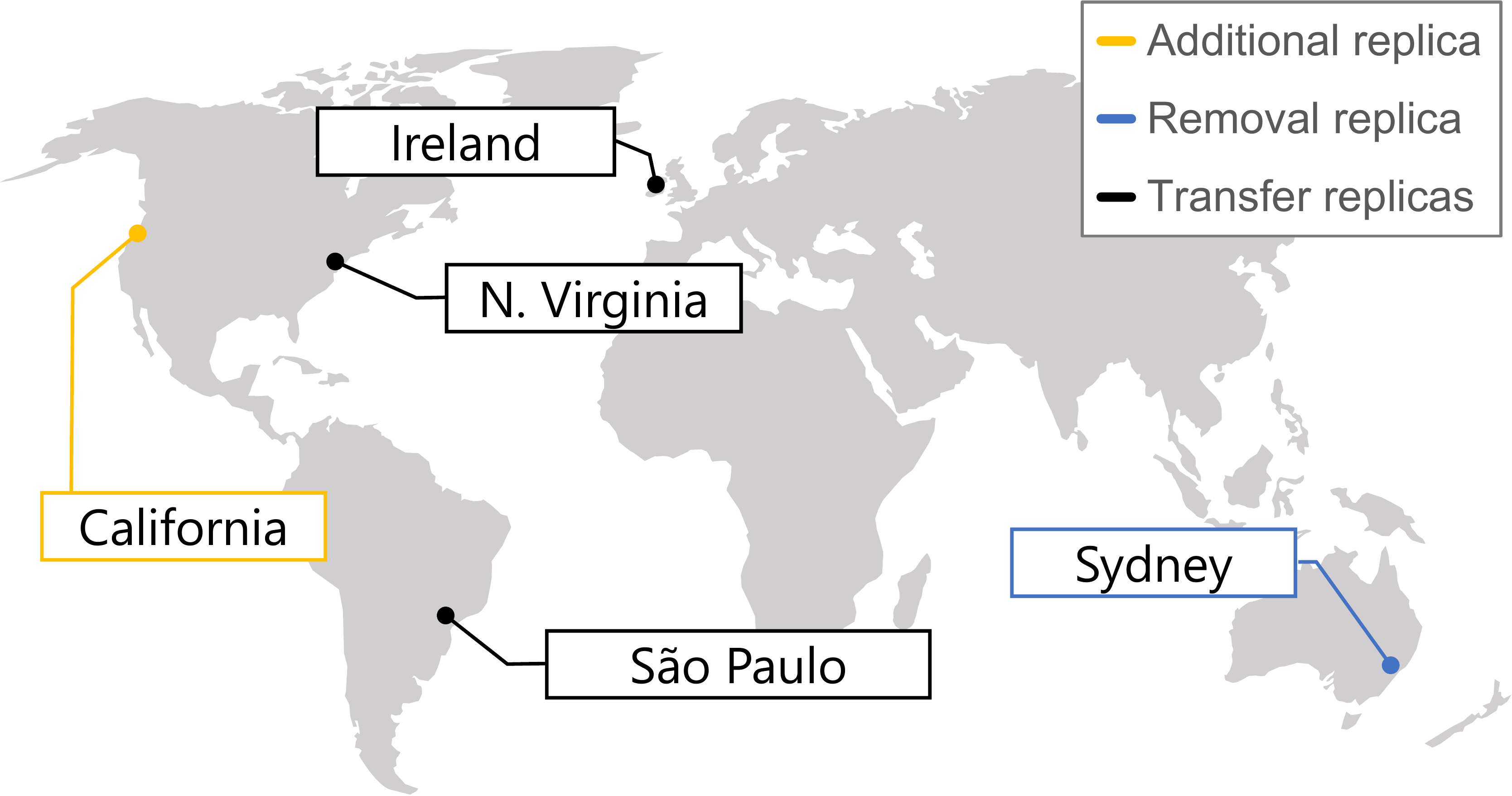}
    \caption{Replica deployment in the experiment of replica migration.}
    \label{fig:relocation-map-california}
\end{figure}

Here, we move a replica with high latency between replicas to another location with low latency by dynamic replica replacement and measure the change in latency.
Unlike Sect.~\ref{sec:dynamic-relocation-when-replica-lantecy-becomes-bad}, the removal replica also participates in the state transfer to the additional replica.
As shown in Fig.~\ref{fig:relocation-map-california}, replicas are placed in five locations: the Worldwide Group (Ireland, North Virginia, Sydney, and São Paulo) and California.
The removal replica is in Sydney, and the additional replica is in California.
The leader replica and a client are located in North Virginia.
The experiment was conducted on February 3, 2022.

\begin{figure}[t]
    \centering
    \includegraphics[width=75mm]{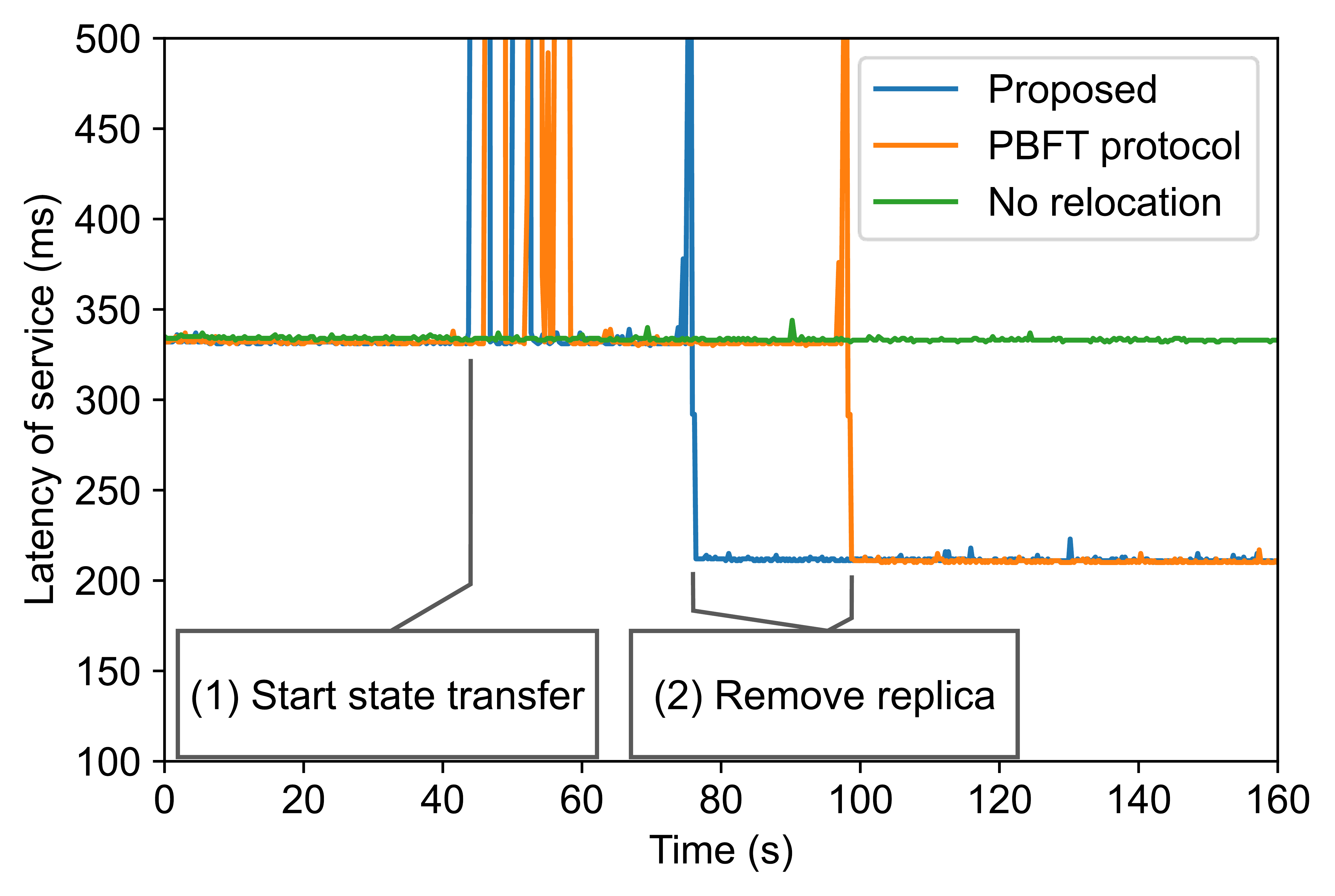} 
    \caption{Latency change in the experiment of replica migration.}
    \label{fig:replica-relocation-sydney-to-california}
\end{figure}

The change in request latency is shown in Fig.~\ref{fig:replica-relocation-sydney-to-california}.
At (1) of Fig.~\ref{fig:replica-relocation-sydney-to-california}, replicas started state transfer to the additional replica in California.
Similar to Fig.~\ref{fig:replica-relocation-london-to-ireland}, the system request processing stopped for approximately one second in the proposed method and PBFT status transfer.
At (2), the state transfer was completed and the additional replica was added to the system, so the removal replica in Sydney was removed from the system.
Compared to Sydney, California has a lower latency to the leader replica; thus, the system latency was improved drastically compared to that before (1).
PBFT took 55 seconds for a state transfer, while the proposed method took 35 seconds.
This result indicates that the proposed method is also effective for the dynamic replacement of a replica with high latency between replicas.

\section{Conclusion}
\label{sec:conclusion}

In this paper, we proposed a state transfer method suitable for geographic SMR.
The proposed method addresses the problems caused by the instability of the communication bandwidth specific to geographic SMR, that is, variation and non-uniformity of the communication bandwidth.
This is achieved by passively estimating the communication bandwidth between replicas and dynamically adjusting the amount of state that each transfer replica sends to a recovery replica according to their communication bandwidth.
The evaluation results showed that the proposed method can deal with the communication bandwidth variations and reduce the state transfer time by up to 47\% compared to the existing method.
Moreover, we showed that the proposed method can be used in dynamic replica replacement, which allows us to improve the service latency of geographic SMR by moving a slow replica to another location.
For future work, we plan to reduce the service outage time while adding or removing replicas and to develop a state transfer method that relocates multiple replicas simultaneously by transferring the state to multiple replicas simultaneously.
This approach would allow us to realize faster replacement.

\section*{Acknowledgment}
This work was supported by JSPS KAKENHI Grant Number JP18K18029.

\end{document}